\pdfoutput=1
\documentclass[iop]{emulateapj}
\usepackage{natbib}
\usepackage{txfonts}
\usepackage{graphicx}
\usepackage{wasysym}
\usepackage{multirow}

\newcommand{\sfrone}{$M_{\astrosun}$~yr$^{-1}$}
\newcommand{\sfrtwo}{$M_{\astrosun}$~yr$^{-1}$~kpc$^{-2}$}
\newcommand*{\vcenteredhbox}[1]{\begingroup
\setbox0=\hbox{#1}\parbox{\wd0}{\box0}\endgroup}

\shortauthors{Watkins et al.}

\begin{document}

\title{Searching for diffuse light in the M96 galaxy group}
\shorttitle{Diffuse light in the M96 Group}

\author{Aaron E. Watkins\altaffilmark{1}, 
  J. Christopher Mihos\altaffilmark{1},
  Paul Harding\altaffilmark{1}, 
  John J. Feldmeier\altaffilmark{2}}

\altaffiltext{1}{Department of Astronomy, Case Western Reserve
  University, Cleveland, OH 44106, USA}
\altaffiltext{2}{Department of Physics and Astronomy, Youngstown 
State University, Youngstown, OH 44555, USA}

\begin{abstract}

We present deep, wide-field imaging of the M96 galaxy group (also
known as the Leo I Group). Down to surface brightness limits of
$\mu_{B}=30.1$ and $\mu_{V}=29.5$, we find no diffuse, large-scale
optical counterpart to the ``Leo Ring", an extended HI ring
surrounding the central elliptical M105 (NGC 3379). However, we do
find a number of extremely low surface-brightness ($\mu_{B} \gtrsim
29$) small-scale streamlike features, possibly tidal in origin, two of
which may be associated with the Ring. In addition we present detailed
surface photometry of each of the group's most massive members --
M105, NGC 3384, M96 (NGC 3368), and M95 (NGC 3351) -- out to large
radius and low surface brightness, where we search for signatures of
interaction and accretion events. We find that the outer isophotes of
both M105 and M95 appear almost completely undisturbed, in contrast to
NGC 3384 which shows a system of diffuse shells indicative of a recent
minor merger. We also find photometric evidence that M96 is accreting
gas from the HI ring, in agreement with HI data. In general, however,
interaction signatures in the M96 Group are extremely subtle for a
group environment, and provide some tension with interaction scenarios
for the formation of the Leo HI Ring.  The lack of a significant
component of diffuse intragroup starlight in the M96 Group is
consistent with its status as a loose galaxy group in which encounters
are relatively mild and infrequent.

\end{abstract}

\keywords{galaxies: elliptical -- galaxies: groups: individual(Leo I)
  -- galaxies: interactions -- galaxies: stellar content -- galaxies:
  structure}

\newpage

\section{Introduction}

In the current $\Lambda {\rm CDM}$ cosmological paradigm, massive
galaxies form hierarchically over time, via the continual accretion of
smaller objects \citep{searle78, davis85, frenk88, wf91, jenkins01,
  springel05}.  Massive galaxies may interact with one another as well,
leaving behind readily observable signatures in the form of tidal
streams or drastic changes in morphology, stellar populations, and
kinematics \citep[e.g.,][]{arp66, toomre72, toomre77, hernquist92,
  barnes92, barnes02}.  Even after the initial construction of
galaxies has ended, however, cosmological simulations indicate that
there should remain a wealth of low-mass satellites that continue to
interact with the parent galaxy over time \citep[e.g.,][]{frenk88,
  gelb94, springel05, springel08}.  Observations of the Local Group
indeed show a number of low-luminosity satellites around the Milky Way
and M31 \citep{hodge71, feitzinger85, mateo98, belokurov06, weisz11,
  mcconnachie12}, lending credence to the idea.  However, important
discrepancies between theory and observation remain, such as the
`missing satellites problem' \citep{klypin99, moore99}, demonstrating
the importance of continuing to test the theory observationally in a
variety of ways.

Probing the low surface brightness outskirts of galaxies via deep
surface photometry provides one such observational test. Under
proposed ``inside-out'' galaxy formation models \citep{matteucci89,
  bullock05, naab06, hopkins09, vandokkum10, kauffmann12}, even
apparently undisturbed galaxies should show signs of recent accretions
or interactions in subtle ways in their outskirts, where restoring
forces are low and dynamical times are long. Morphological signatures
of accretion such as tidal features, warps, and disk asymmetries
\citep{toomre72, bullock05, martdelg10} may trace the record of past
encounters, while changes in the structural properties and stellar
populations of the outer disk can be probed via quantitative
photometry and color profiles \citep[e.g.][]{vanderkruit79, pohlen02,
  trujillo09, mihos13a}.  Finding and categorizing such signatures
thus becomes an important step in constraining theories of galaxy
formation.

Unfortunately, the extended outskirts of galaxies are extremely dim
and diffuse.  However, with the advent of deep wide-field imaging
techniques, over the past decade a wide variety of faint tidal
features have been identified around nearby galaxies.  These features
span a wide range of morphologies, from loops, plumes, and shells
\citep{forbes03, martdelg08, martdelg09, martdelg10, atkinson13,
  mihos13a} to extended streams, diffuse stellar halos, and
large-scale intracluster light \citep{uson91, scheick94, gonzalez00,
  feldmeier04a, adami05, mihos05, rudick10b}.  Gaseous accretion
events also appear to influence the properties of disks
\citep{sancisi08}, fueling new star formation in their low density
outskirts and driving the formation of extended ultraviolet (XUV)
disks \citep{thilker05, gildepaz05, thilker07, lemonias11}. This
extended star formation will in turn affect the age and metallicity
distributions, and hence the colors, of the underlying stellar
populations in these regions. Meanwhile, internal processes within
disks may scatter stars outwards, again resulting in changes in the
structure and stellar content of the outer disk \citep{sellwood02,
  debattista06, roskar08a, roskar08b}.

Local environment also must play a role in the accretion history of
galaxies and the evolution of their outer regions. In massive
clusters, for example, frequent and rapid interactions between
galaxies tend to produce plumes and streams of starlight, which are
shredded over time by interactions within the cluster potential to
form the more diffuse intracluster light (ICL) seen in clusters such
as Virgo \citep{rudick09}. In group environments, where the velocity
dispersions are generally lower, slow encounters lead to strong
dynamical friction and rapid merging \citep{hickson77, barnes89,
  taranu13}.  Contrary to cluster environments, then, galaxies in
groups may exhibit long-lived accretion shells or tidal tails in their
outskirts \citep[e.g.,][]{quinn84, hernquinn87, law05}.  Meanwhile,
the low density environment around isolated field galaxies should lead
to fewer recent interactions and accentuate secular evolution
processes that are internal to galaxies \citep[e.g.][]{kraljic12}. A
complete survey of the processes that shape galaxies thus requires an
investigation across all environments.

Of particular interest is the group environment, the most common
environment for galaxies to reside. Intragroup starlight (IGL) in
loose groups\footnote{Because of their irregular nature, a
    precise definition of a `loose group' is elusive.  However,
    characteristic properties are often adopted to classify them, such
    as a small number of luminous galaxies within a 0.5 Mpc radius and
    a velocity spread of $\sim$350 km/s \citep[see, e.g.,][]{devau75,
      geller83, maia98}.  Leo I has been classified as a group using
    these criteria by \citet{devau75}, \citet{huchra82}, and
    \citet{tully87}.} is of significant interest across many studies
  of galaxy and galaxy cluster evolution. The fraction of baryons
  within galaxy groups and clusters constrains models of star
  formation and cluster assembly, and there is strong evidence that
  the amount of intragroup and intracluster starlight may be a
  significant contributor to the baryon fraction of these systems
  \citep[e.g.][]{gonzalez13, budzynski14}. The loose group environment
  may also dynamically ``preprocess'' galaxies that later fall into
  galaxy clusters and produce ICL \citep{mihos04}, and substructure
  within the diffuse IGL should trace this process in dynamically
  evolving galaxy groups.

However, the mean fraction of stars in an IGL component, and how that
fraction varies with group/cluster mass, is substantially uncertain.
Theoretical studies give conflicting results: some studies have found
relatively constant mean IGL/ICL fractions with group/cluster mass,
but with large intrinsic scatter \citep[e.g.][]{sommer05, sommer06,
  monaco06, henriques10, puchwein10, guo11, contini14}.  Other
theoretical studies have found that the mean IGL/ICL fraction
(sometimes generalized as the intrahalo stellar fraction; IHL)
systematically increases with group/cluster mass
\citep[e.g.][]{murante07, purcell08, watson12}. 

Observationally, wide-field imaging studies of 0.01~$< z <$~0.4 galaxy
clusters over a large range of cluster mass have also found different
behaviors for the IGL/ICL fraction as a function of cluster mass
\citep[e.g.][]{lin04, zibetti05, gonzalez07, gonzalez13,
  budzynski14}. Studies of nearby individual loose galaxy groups
searching for individual intragroup stars also have mixed results.
Observational searches for intragroup red giant stars and intragroup
planetary nebulae have generally found very small fractions of IGL
\citep[$<$ 2\%; e.g.][]{castro03, feldmeier04b, durrell04}.  Usually,
these surveys only observe a portion of the galaxy group, and if the
IGL has an uneven spatial distribution \citep[e.g.][]{sommer06}, the
IGL fractions found may not be reflective of the entire group.
However, in stark contrast, \citet{mcgee10} have estimated that
$47$\%$^{+16}_{-15}$ of the stellar mass in galaxy groups is in the
IGL component through detections of spectroscopically confirmed
hostless intragroup SNe Ia that were within the Sloan Digital Sky
Survey \citep[SDSS;][]{york00} SN survey.

\begin{figure*}
  \centering
  \includegraphics[scale=0.3]{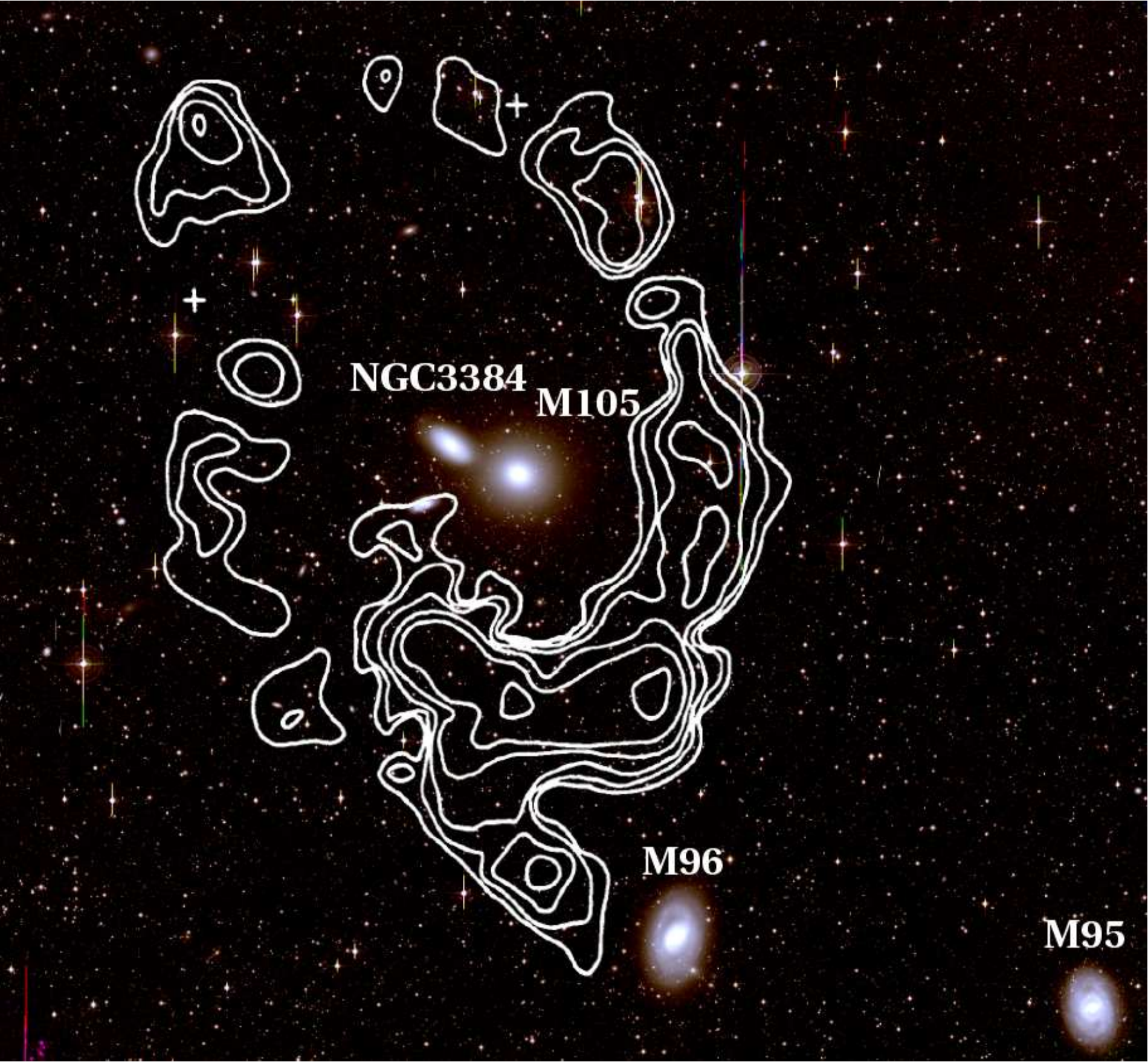}
  \caption{False-color image of the M96 Group, made from a composite
    of our B and V mosaics.  The four largest group members are
    labeled; the small spiral galaxy south of NGC 3384 is NGC 3389 and
    is most likely part of a separate group called Leo II, roughly 10
    Mpc behind the M96 Group \citep[and references
      therein]{stierwalt09}.  HI contours from \citet{schneider89} are
    overlaid in white.  The image is $1.8^{\circ} \times 1.8^{\circ}$
    ($350 \times 350$ kpc$^{2}$).  North is up and east is to the left.
        \label{fig:leopretty}}
\end{figure*}

To improve this situation, more searches for IGL need to be undertaken
in galaxy groups over a large range of mass. The M96 Group (sometimes
referred to as the Leo I Group), shown in Figure \ref{fig:leopretty},
helps serve this purpose well. It is the nearest loose group\footnote{We
adopt a distance of 11 Mpc for the M96 Group \citep{graham97}; at this
distance, 1\arcmin\ corresponds to 3.2 kpc.} that contains both early
and late type massive galaxies \citep{devau75}, and hosts a rather
unusual feature -- a roughly 200 kiloparsec diameter broken ring of
neutral hydrogen, first discovered by \citet{schneider83}. This ring is
fairly diffuse; from Arecibo data, column densities range from $6.4
\times 10^{19}$ cm$^{-2}$ to as low as $2 \times 10^{18}$ cm$^{-2}$
\citep{schneideretal89}, although smaller clumps with densities of
roughly $4 \times 10^{20}$ cm$^{-2}$ were also detected using the Very
Large Array \citep{schneider86}. These higher density clumps appear to
be confined to the southernmost portion of the ring, and include a
bridge-like feature apparently connecting the ring to M96 (NGC 3368),
the most massive spiral galaxy in the group. Two other galaxies -- the
E1 galaxy M105 (NGC 3379) and the S0 galaxy NGC 3384 -- appear to lie
near the center of the ring, implying some relation between these three
galaxies and the intergalactic HI. A fourth galaxy, the barred spiral
M95 (NGC 3351), is associated with the group as well, but due to its
relatively large projected separation may not be associated with the
ring complex itself. The group also lies near in the sky
($\sim$8$^{\circ}$ away) and at a somewhat similar distance to the M66
Group, leading to speculation that the two groups are actually each part
of a larger complex of galaxies \citep{stierwalt09}.

The origin of the Leo HI Ring is still under debate.  Kinematics
gleaned from the HI observations show that it appears to be
well-modeled by an elliptical orbit with a focus at the
luminosity-weighted centroid of NGC 3384 and M105
\citep{schneider85}. The rotational period of this model orbit is
approximately 4 Gyr, and if the Ring is primordial, this period
presents a rough lower limit on its age \citep[see][]{schneider85}.
The apparent lack of associated starlight and star formation supports
this primordial origin hypothesis \citep{pierce85, schneideretal89,
  donahue95}, as does the similarity between the radial velocities of
the Ring and the galaxies within the group, both of which trend higher
in the northwest and lower in the southeast \citep{schneider85}.
Additionally, \citet{silchenko03} argue that circumnuclear stellar and
gaseous disks in NGC 3384, M96, and M105 show ages and kinematics
consistent with accretion from the Ring taking place on gigayear
timescales, which may rule out a more recent formation scenario.  Of
course, crossing times in the group are much shorter than 4 Gyr
\citep{pierce85}, which poses problems for the stability of the ring
-- maintaining its shape would require a shepherding mechanism,
presumably by M96 \citep{schneider85} -- and the only galaxy in the
group whose disk appears aligned with the Ring is NGC 3384, unusual if
all of the galaxies formed from a single rotating cloud.

An alternative hypothesis was proposed by \citet{rood84}, involving
the collision of two spiral galaxies.  Under this model, the two
galaxies collided nearly head-on, stripping most of the gas from one
of the galaxies and producing an expanding density wave much like that
proposed to explain the morphology of ring galaxies
\citep[e.g.,][]{lynds76, gerber92, mazzei95, berentzen03}. A more
recent simulation by \citet{md10} managed to reproduce qualitatively
all of the important features of the Ring (the annular shape, the
apparent rotation, the bridge-like feature described above, and the
lack of an apparent visible light counterpart).  The simulation
collided two gaseous disk galaxies, one of which transformed into an
S0: a good facsimile of M96 and NGC 3384.  In yet another scenario,
the total disruption of a gas-rich, low surface brightness (LSB)
galaxy would also leave behind an HI ring with the LSB galaxy's
stellar disk completely unaffected \citep{bekki05}.  In general, the
main problem to solve in a collisional hypothesis is to reproduce the
apparent lack of optical light associated with the ring, as slow
encounters in the loose group environment should be effective at
liberating stellar material from the galaxies as well. 

To date, neither the primordial origin nor the collisional hypothesis
for forming the Leo Ring has been confirmed. No extended diffuse
starlight has been found associated with the HI ring; the only known
stellar counterpart to the Ring is that discovered by \citet{thilker09}
in the form of three small far- and near-ultraviolet sources located
within the ring's highest density HI complexes \citep[as measured
by][]{schneider86}. An optical counterpart to at least one of these
clumps was later found by \citet{md10}, with a measured color of $g' -
r' = 0.2$, indicative again of a young stellar population. The presence
of these clumps thus raises the intriguing possibility that these stars
represent the youngest (and therefore brightest) component of an
underlying (and as-yet undetected) diffuse stellar counterpart to the
ring. All of this makes the M96 Group a particularly intriguing target
in a search for the low surface brightness signatures of historical
encounters.

This paper presents deep surface photometry of the M96 Group,
conducted using Case Western Reserve University's Burrell Schmidt
telescope on Kitt Peak and taken as part of an ongoing project to
study the outskirts of nearby galaxies across a range of
environments. Our deep imaging, covering approximately nine square
degrees in two filters, probes the diffuse outskirts of the group
members and the Leo Ring down to a surface brightness of $\mu_V \sim
29.5$, with reliable $B-V$ colors down to $\mu_V \sim 28.0$. We use
the data to search for diffuse light in the Leo Ring that may trace
any past interaction events, as well as to study the structure and
stellar populations in the outer disks and halos of the group
galaxies.

\section{Observations and Data Reduction}

The full description of our observational and data reduction
techniques can be found in \citet{rudick10a} and \citet{mihos13a},
however, for the sake of clarity, we describe the most important
details here, along with any notable changes to the procedure.

\subsection{Observations}

We observed the M96 Group using the 0.6/0.9m CWRU Burrell Schmidt
telescope located at Kitt Peak National Observatory. The telescope has
a $1.65^{\circ} \times 1.65^{\circ}$ field of view and images onto a
$4096 \times 4096$ STA0500A CCD\footnote{The CCD was backside
  processed at the University of Arizona Imaging Technology
  Laboratory.} for a pixel scale of $1.45$\arcsec\ pixel$^{-1}$.  We
observed in two filters: a modified Johnson B (Spring 2012) and
Washington M (Spring 2013). We used Washington M as a substitute for
Johnson V, as it covers a similar bandpass but avoids night sky
airglow emission lines \citep{feldmeier02}; thus, as Washington M is
slightly bluer than Johnson V, we used a modified B filter which is
200 \AA\ bluer than standard Johnson B in order to maintain a
comparable spectral baseline. For the actual analysis, we converted
our magnitudes to standard Johnson B and V using observations of
\citet{landolt92} standard fields.  We took data only on moonless
photometric nights, although it should be noted that Mars was less
than $3^{\circ}$ from our target fields during part of the Spring 2012
run and hence may have contributed some low-level scattered light.

Along with target pointings, we also observed offset sky pointings to
construct a night sky flat. To minimize uncertainties in the final
flat due to changes in sky conditions and telescope flexure, we kept
the telescope oriented at roughly the same pointing for flat- field
and object exposures by alternating between the two. We dithered
individual target exposures randomly by up to half a degree to
increase the imaging area and reduce artifacts due to scattered light
and large-scale flat fielding errors. Exposure times were 1200 seconds
in B (resulting in sky levels of 700 -- 900 ADU pixel$^{-1}$) and 900
seconds in M (1200 -- 1400 ADU pixel$^{-1}$). Our final B and V
mosaics contain 41 exposures each, with roughly 30 accompanying blank
sky frames per band taken between 0.25 -- 1 hr in right ascension and
1 -- 4$^{\circ}$ in declination from the M96 Group. However, we note
that the final flat field was produced using blank skies near every
object we imaged during the course of this project, for a total of
$\sim$100 frames in B and $\sim$120 in V (discussed in more detail
below). Our total on-target exposure time in the M96 Group is 13.7
hours in B and 10.25 hours in V, although the time spent in each part
of the mosaic varies due to dithering across the group's large angular
size ($\sim 3^{\circ} \times 3^{\circ}$).

\subsection{Data Reduction}
For each frame, we applied an overscan correction and subtracted a
nightly mean bias frame using IRAF\footnote{IRAF is distributed by the
  National Optical Astronomy Observatory, which is operated by the
  Association of Universities for Research in Astronomy (AURA) under
  cooperative agreement with the National Science Foundation.}.  We
applied no crosstalk corrections; close examination of the images
revealed that any visible crosstalk was at such low levels, such a
correction was unnecessary.  We also corrected each year's data for
nonlinear chip response in each quadrant of the CCD using a 4$^{th}$
order polynomial correction.  Finally, we applied a world coordinate
system to each image to construct the final mosaic.

After a preliminary flat-field correction, we derived a photometric
zeropoint for each target frame (directly correcting for airmass
effects) using SDSS stars as standards, converted from \emph{ugriz} to
Johnson B and V using the conversion formula given by
\citet{lupton05}. We excluded stars outside of the color range $B-V =
0$ to $1.5$ from the fit \citep{ivezic07}. To derive each filter's
color term, we took exposures of Landolt standard star fields
\citep{landolt92} at varying exposure times at the beginning and end
of each night, as using Landolt stars avoids any error inherent in a
color conversion formula. From our final mosaics, we were able to
recover converted SDSS magnitudes to $\sigma_{V}=0.03$ mag and B-V
colors to $\sigma_{B-V}=0.05$, using SDSS stars in the magnitude range
$V=15$-$17$ and the color range $B-V=0.0$-$1.5$.

After applying zeropoints, and assuming that the night sky has an
average B-V color near unity \citep{krisciunas97, patat03}, we
measured sky brightnesses from $\mu_{B} = 22.32$ to $21.89$ and from
$\mu_{V} = 21.72$ to $21.43$.  These values changed throughout the
course of a night, as well as from night to night due to changes in
solar output, lunar phase, and atmospheric conditions (e.g. dust,
humidity, etc.).  However, the faintest sky brightness measurement can
act as a useful secondary check, by comparing our measured sky
brightnesses to previously measured values under similar conditions.
To measure the actual sky brightness, we first removed our airmass
corrections of 0.27 mag and 0.16 mag in B and V, respectively, since
the source of the sky brightness is mostly within the Earth's
atmosphere.  We then compared our faintest values ($\mu_{V} = 21.88$,
$\mu_{B} = 22.59$) to the empirical relation of \citet[see their
  equation 5]{krisciunas97}, which correlates the average observed
night sky brightness to the 10.7cm solar flux.  Taking the monthly
10.7cm values from the Dominion Radio Astrophysical
Observatory\footnote{data located at:
  http://www.spaceweather.ca/solarflux/sx-eng.php}, we find an
adjusted expected sky brightness of $\mu_{V} = 21.75 \pm 0.06$ and
$\mu_{B} = 22.65 \pm 0.06$.  Given the uncertainties in this
calculation, this shows that our overall results are stable and
robust.

\begin{deluxetable*}{l c c c c c c c c c c}
\tabletypesize{\footnotesize}
\tablecaption{Properties of streamlike features \label{table:streams}}
\tablecolumns{11}
\tablehead{
\colhead{}\vspace{-0.15cm} & \colhead{Dimens} & \colhead{Dimens} & & & & & & & &\\
\colhead{Region}\vspace{-0.15cm} & & & \colhead{$\mu_{B}$} & \colhead{$\mu_{V}$} & \colhead{$M_{B}$} &
\colhead{$M_{V}$} & \colhead{B-V (Range)} & \colhead{$\mu_{B,lim}$}
& \colhead{$\mu_{V,lim}$} & \colhead{$10^{6}L_{\astrosun,B}$} \\
\colhead{} & \colhead{(\arcmin)} & \colhead{(kpc)} & \colhead{} & \colhead{} & \colhead{} & \colhead{}
 & \colhead{} & \colhead{} & \colhead{} & \colhead{}}
\startdata
A & $1.3 \times 9.4$ & 4 $\times$ 30 & 29.3$\pm$0.3 & 28.8$\pm$0.3 & 
-12.2 & -12.6 & 0.45(-0.14 -- 1.04) & 30.9 & 30.5 &  12\\
B & $1.1 \times 5.2$ & 3.5 $\times$ 17 & 29.2$\pm$0.2 & 28.6$\pm$0.3 &
 -11.6 & -12.1 & 0.56(-0.08 -- 1.15) & 30.9 & 29.9 & 7\\
C North & $1.1 \times 9.6$ & 3.5 $\times$ 31 & 29.8$\pm$1.0 &
 29.6$\pm$0.8 & -11.7 & -11.9 & 0.21(-1.60 -- 2.55) & 30.1 & 29.7 & 7\\
C South & $1.3 \times 7.0$ & 6 $\times$ 22 & 29.9$\pm$1.0 &
 29.1$\pm$0.8 & -11.4 & -12.1 & 0.77(-1.04 -- 3.11) & 30.1 & 29.7 & 6\\
C Total & $1.2 \times 60.1$ & 3.5 $\times$ 200 & 29.9$\pm$1.0 &
 29.4$\pm$0.8 & -13.9 & -14.4 & 0.49(-1.32 -- 2.83) & 30.1 & 29.7 & 56$^{a}$\\
\enddata
\tablecomments{$a$: Assumes a continuous structure from C North to C
  South: see text.}
\end{deluxetable*}

Accurate flat-fielding is paramount for reliable surface photometry of
low surface brightness features. As mentioned previously, we used
blank sky fields near our target objects to construct a master night
sky flat.  For each sky frame, we masked stars and background objects
by running IRAF's \emph{objmask} task twice -- once to generate a
basic mask, then again with the first mask applied to mask even
fainter sources; both times we masked objects using a threshold of
  4 $\sigma$ above background and 5 $sigma$ below \citep[more details
    on the masking procedure can be found in][]{rudick10a}.  We then
calculated the median values in $32 \times 32$ pixel blocks, and
binned the images using these median values in order to enhance any
faint, extended contaminants that were missed by \emph{objmask}, such
as flares from stars off the edge of the chip, extended wings of
stars, and internal reflections. These features we masked by hand, and
images with obvious large-scale contamination (faint cirrus clouds,
unusual electronic noise) were discarded. We median combined all of
these masked sky images to create a preliminary sky flat, re-flattened
the sky frames using this flat, and modeled and subtracted sky planes
from each flattened image. We used these flattened, sky-subtracted
images to create a new flat, which we then used as a basis to repeat
the process. The flat-field converged after five such iterations,
resulting in the final generation master sky flat.

To test for systematics in the flat-fielding, we used this process to
construct three types of master flats: master flats by object (all
objects imaged during the course of this project, including targets
other than the M96 Group), master flats by observing run, and a master
flat made from all skies taken over the course of an observing season.
We divided all pairs of these master flats to check for systematic
differences, and we determined that there only existed a low-level
(0.2\%) residual plane that varied from run to run. As such, we
created flat fields for each observing run by using the master flat
generated from all sky images, corrected for each run's residual
plane. In this way, our final B flat combined 98 total sky images, and
our final V flat combined 123 total sky images.  We note that the
consistency of the flats (including even those taken nearer to Mars
than the M96 Group) shows that scattered light from Mars was not a
problem in constructing the flat field.

Bright stars can also introduce extended halos of light due to
reflections between the CCD, dewar window, and filter; we modeled and
removed these using the process described in \citet{slater09}, for
which we give a brief summary here. To generate visible reflections,
we took long (1200 s and 900 s in B and M, respectively) exposures of
zeroth magnitude stars (Procyon in B and Regulus in M). We modeled
these reflections as annuli of constant brightnesses scaled by the
total brightness of the parent star. We also measured and modeled the
offset of each reflection from the parent star as a function of the
star's position on the CCD. Lastly, we modeled the star's point spread
function out to where the flux from the star drops to 0.3 ADU
($\mu_{B} \sim 31$). We then subtracted these models from each bright
star in each image. Bright stars also produce visible off-axis
reflections colloquially known as ``Schmidt Ghosts'' \citep{yang02};
we masked these as well for stars brighter than $V=10.5$. To obtain
accurate photometry for the bright stars producing these various
reflections, we used short exposures of our target fields; for the
brightest stars we used magnitudes given in the Tycho II catalog
\citep{hog00}, and in a few cases (for stars with $m_{B} \lesssim 8$)
we had to manually adjust the magnitudes to achieve the best
subtraction.

We then masked and median binned the images into $32 \times 32$ pixel
blocks, hand-masked any scattered light pattern in each object frame
in the same manner as the blank skies, as well as any persistent
bright objects in each frame. We fitted a sky plane to each of these
binned, masked images, and subtracted this sky off of the original
images.  Finally, we mosaiced all of the resulting images using IRAF's
\emph{Wregister} and \emph{imcombine} tasks, scaling each image to a
common zeropoint, and creating $9 \times 9$ pixel and $18 \times 18$
pixel masked and median binned images to help identify faint features
and improve the per pixel statistics in the final mosaics\footnote{For
  the sake of efficiency, for the remainder of the paper we will refer
  to the process of masking and calculating the median in blocks as
  simply `binning'. As such, references to e.g. the `$9 \times 9$
  binned image' refer to the image that has been masked and then
  median-binned into $9 \times 9$ pixel blocks.}.  The masking
  process used to generate these images typically only masks pixels
  brighter than $\mu_{B} \approx 27$ and $\mu_{V} \approx 26$, so that
  the faint, diffuse signals we are searching for remain visible.

\section{Large scale diffuse light}

\begin{figure*}
  \epsscale{0.58}
  \plotone{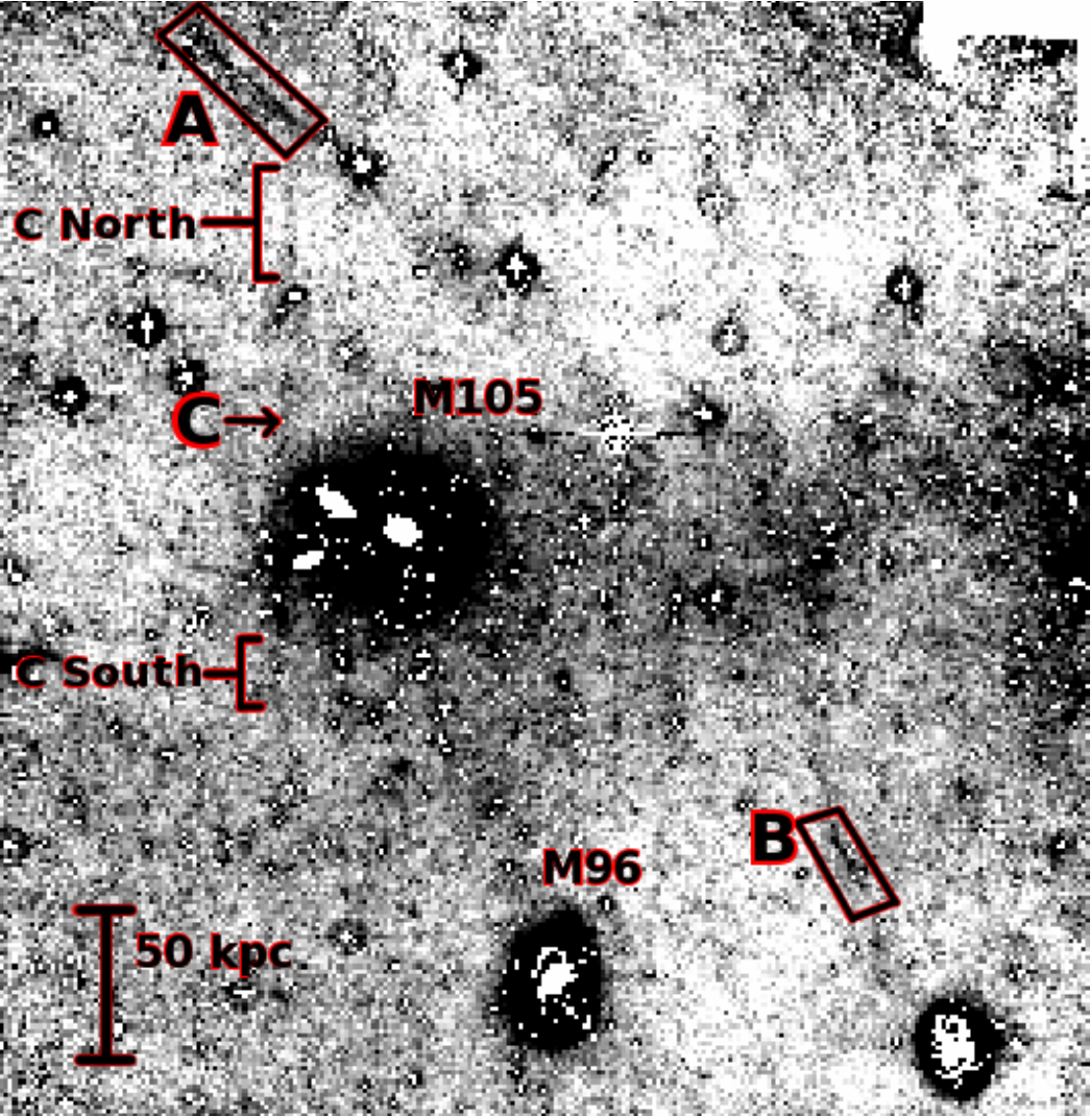}
  \plotone{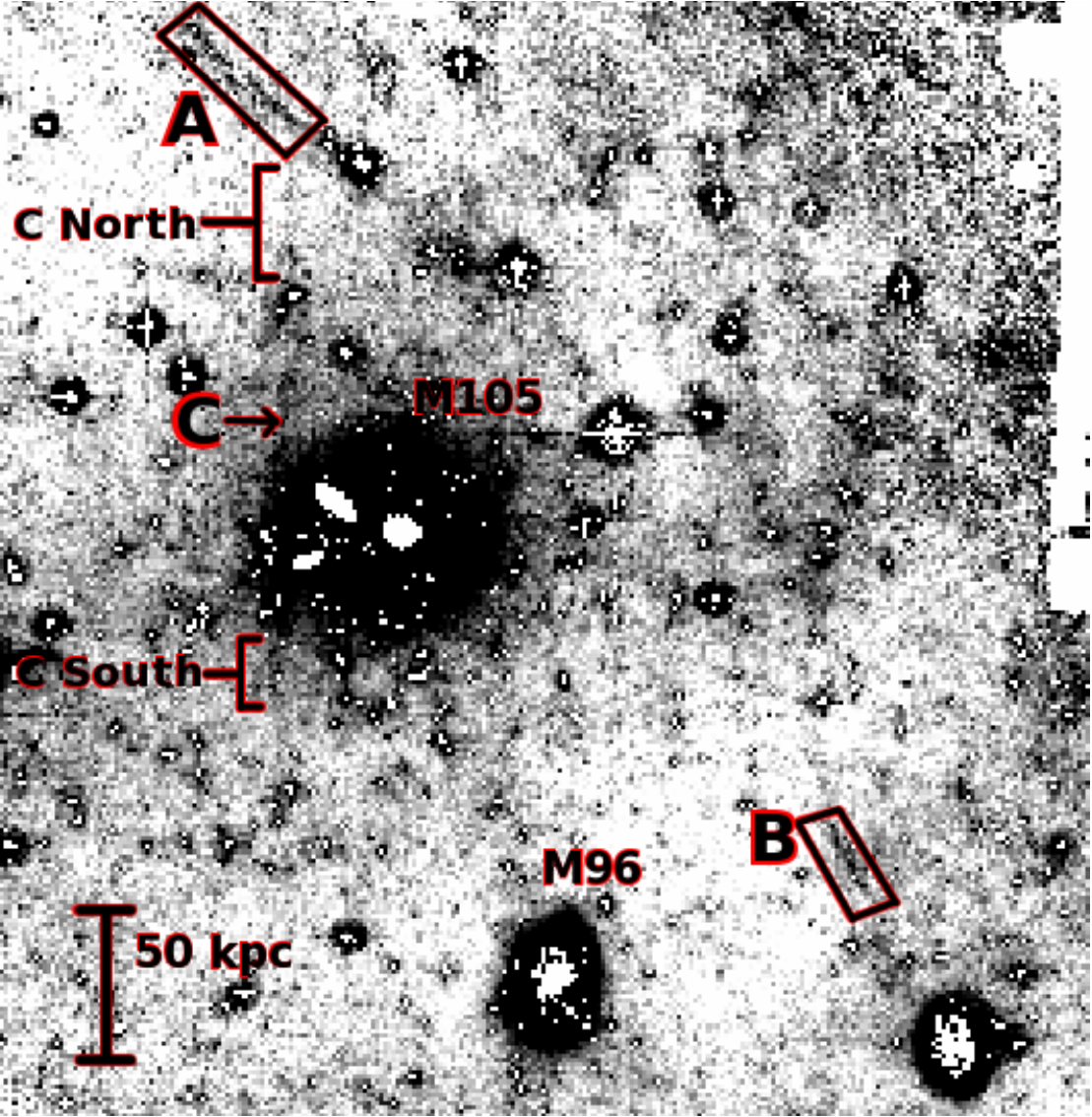}
  \caption{Our final B mosaic (left) and V mosaic (right), with pixels
    binned $18 \times 18$ to improve the clarity of the features of
    interest.  The three regions discussed in the text are labeled A,
    B, and C; due to the faintness of region C, its general position
    is indicated only by an arrow.  C North and C South indicate where
    the photometry for the feature was performed, with the brackets
    showing the extent of the regions sampled.  M105 and M96 are
    labeled for reference.  In both images, dimensions are
    $1.9^{\circ} \times 1.9^{\circ}$ ($360 \times 360$ kpc$^{2}$), and
    north is up and east is to the left.  Boxes are for illustration
    purposes only.  Contrast is such that pixels saturate below $\mu
    \sim 28.5$.  Images are masked at high ($\mu \lesssim 24$) surface
    brightness.
       \label{fig:mosaic}}
\end{figure*}

\begin{figure*}
  \epsscale{0.97}
  \plotone{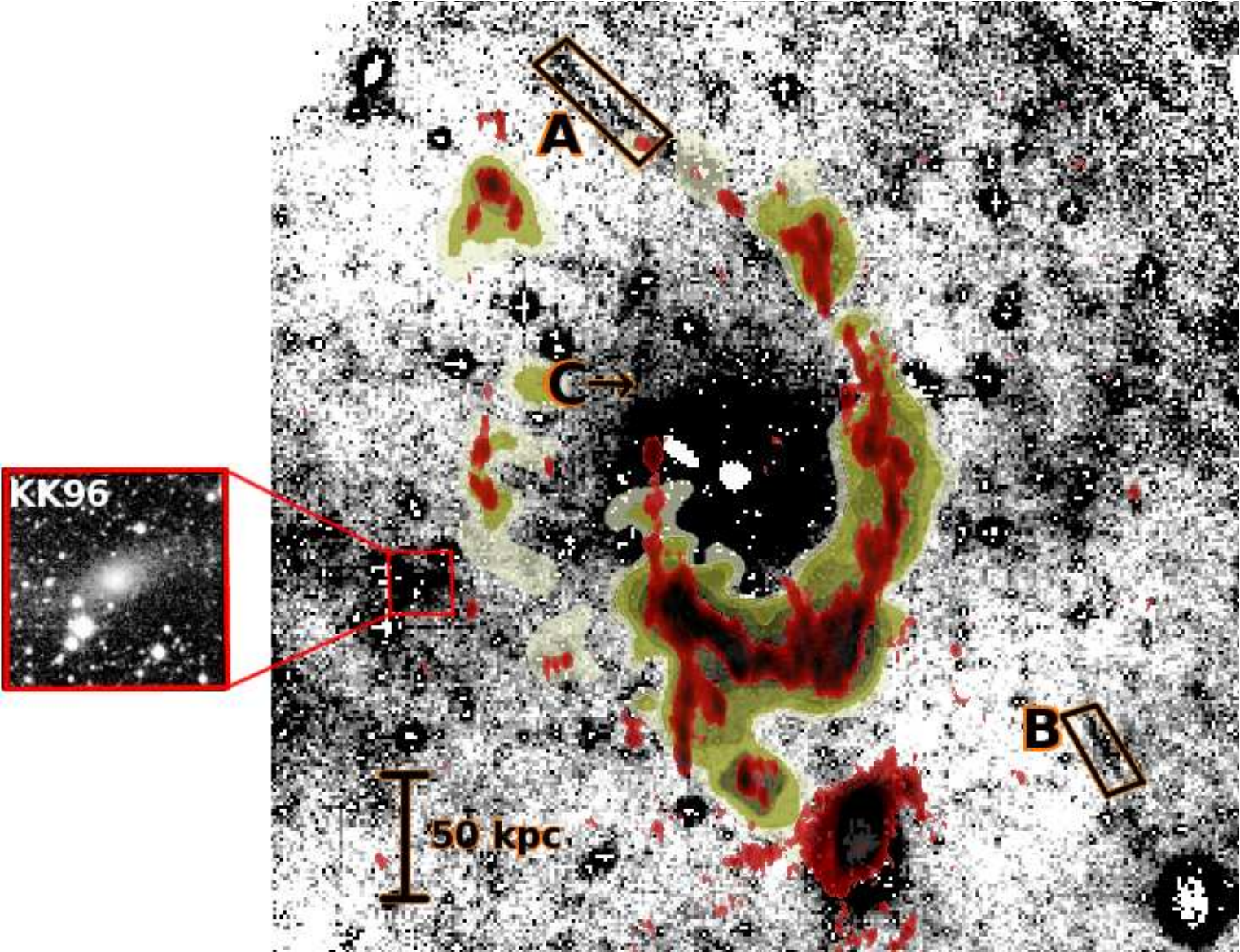} \hspace*{3cm}
  \caption{Our final V mosaic, binned $18 \times 18$ as in Figure
    \ref{fig:mosaic}, with HI contours from \citet{schneider89} and
    \citet{md10} overlaid in yellow and red, respectively. The three
    regions from Figure \ref{fig:mosaic} are labeled for reference, as
    is the dwarf galaxy KK96 (see text).  The inset image is the
    unbinned image of KK96, zoomed in for clarity, to show the tidal
    tails more clearly.  Again, the image is $1.9^{\circ} \times
    1.9^{\circ}$ ($360 \times 360$ kpc$^{2}$) with north up and east
    to the left, and boxes are for illustration purposes only.
        \label{fig:overlay}}
\end{figure*}

\begin{figure*}
  \epsscale{0.85}
  \plotone{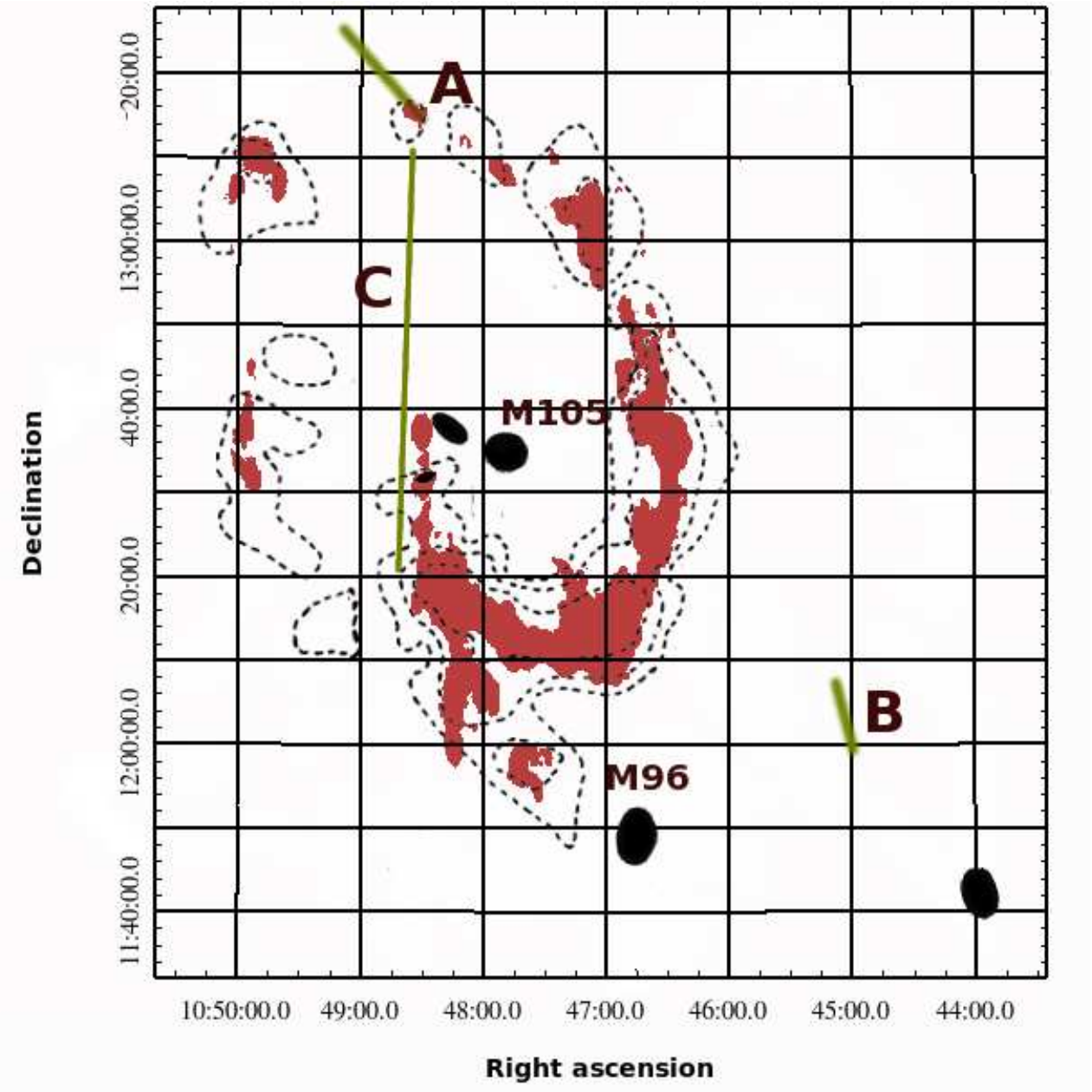}
  \caption{Schematic of the three features in relation to the group
    galaxies and the HI ring, for clarity.
        \label{fig:schematic}}
\end{figure*}

Figure \ref{fig:mosaic} shows our final 18 $\times$ 18 pixel
(26\arcsec $\times$ 26\arcsec, or $1.4 \times 1.4$ kpc$^{2}$) binned B
and V mosaics.  This binning choice is simply to enhance the clarity
of the features of interest; all analyses of the images used either
the unbinned image or a 9 $\times$ 9 binned image.  A schematic is
provided in Figure \ref{fig:schematic} to guide the eye.

The first thing to note in these images is that we see no large-scale
diffuse stellar counterpart to the HI ring. While there is a band of
diffuse light running across the image in the B mosaic, the same feature
does not appear in the V mosaic, which shows a different pattern of
diffuse light elsewhere in the image. Permanent features such as
starlight or foreground Galactic dust lanes (`cirrus') should remain
fixed in position and morphology from image to image, whereas scattered
light patterns will change with respect to the position of the source.
As noted in Section 2.1, Mars was less than $3^{\circ}$ from our fields
in Spring of 2012, when we observed in the B filter, and hence may have
contributed to the scattered light in the final B mosaic. It is less
clear where scattered light may have originated in the V mosaic, with
the brightest nearby source (Regulus) being $\sim 9^{\circ}$ to the
west. Regardless, as the only diffuse signal we see appears to be
scattered light, we find no evidence of actual diffuse starlight in the
Leo HI ring. We place upper limits of $\mu_{B} < 30.1$ and $\mu_{V} <
29.5$ on the surface brightness of any such component, as these are the
average surface brightnesses of the scattered light signals found in the
southern portion of the Ring (where the bulk of the HI mass is found).
We note that this is consistent with an estimate by \citet{castro03} of
$\mu_{B}<32.8$ based on the apparent lack of planetary nebulae
associated with the cloud.

Despite this lack of large scale diffuse light in the Ring, we do find
three smaller features, which are labeled in Figures \ref{fig:mosaic}
and \ref{fig:overlay} and shown schematically in Figure
\ref{fig:schematic}: a streamlike object north of NGC 3384 (A), a
shorter streamlike object northeast of M95 (B), and a nearly vertical
streamlike object west of NGC 3384 and the background galaxy NGC 3389
(C, most clearly visible in the V band image). While these features
are consistent in morphology and position in both mosaics, they are
extremely faint, so to further test their consistency we created three
sets of new B and V mosaics using randomly selected half-samples of
the object frames used to build the final mosaics in each band. All
three objects persisted even in these half-split mosaics. We note also
that all three features are visible in a preliminary data set taken in
2010 in the M band, so the three features appear to be robust. Other
structures are visible in Figures \ref{fig:mosaic} and
\ref{fig:overlay}, such as a V-shaped structure on the images' west
sides, but these were not consistent in morphology amongst all
half-splits and so were not considered for analysis.

Finally, as scattered light from galactic dust (the so-called ``Galactic
cirrus'') can often show similar diffuse morphology, we cross-checked
our mosaics against the IRIS 100 $\mu$m map \citep{miville05} for the
region. As cool dust emits thermally in the far-infrared, any
contamination from galactic cirrus should show corresponding emission in
the 100 $\mu$m map. However, this map shows that the M96 Group lies in a
region with relatively little such foreground pollution, and shows no
spatially corresponding linear structures. The Planck all-sky thermal
dust emission map \citep{planck13}, though lower resolution, confirms
this as well. This lack of detection in the far infrared makes it
unlikely that these features are merely foreground dust, and thus are
likely to be tidal structures located within the M96 Group.

In addition to the linear streams, we also detect diffuse starlight around 
the galaxy KK96 (located southwest of NGC 3384). This galaxy, 
classified as a low surface brightness dwarf spheroidal (dSph) by 
\citet{karachentseva98} and \citet{karachentsev13}, displays clear tidal
tails toward the east and west in our images (clearly visible even in
the unbinned images: see Figure \ref{fig:overlay}), indicating a recent
interaction.  Due to a lack of distance constraints for this galaxy,
its relation to the M96 Group is unclear.  However, aside from the
M96 Group, it appears to have no obvious nearby neighbors that it may
have interacted with.  

We can attempt to constrain KK96's location to some degree through its
optical properties. The galaxy has an apparent magnitude of $V =
16.9$, including all light out to the Holmberg radius
\citep[0.6\arcmin; see][]{karachentsev13} and excluding the faint
tails. We also measure an effective surface brightness
($r_{e}=0.3$\arcmin) of $\mu_{e,V}=25.8$ and central surface
brightness of $\mu_{V}\approx23$. If placed at the distance of the M96
Group, the galaxy would have an absolute magnitude of $M_{V} = -13.3$,
similar to the Fornax dwarf \citep{vandenbergh08}, and an effective
radius of 960 pc, consistent with the empirical $M_{V}$--$\mu_{e,V}$
and $M_{V}$--$r_{e}$ relations for Sph type galaxies from
\citep{kormendy12}. If the system is significantly closer, for example
within the Local Group at 1 Mpc, the galaxy would be underluminous by
$\approx$ two magnitudes for its surface brightness, given the
Kormendy \& Bender relation. However, given the scatter in the dSph
scaling relationships, the system is also consistent with being a
member of the background Leo II galaxy group \citep[at a distance of
  $\sim$20 Mpc; see][]{stierwalt09}. So while it is {\it plausible}
that the system is being tidally stripped by its interactions within
the M96 Group, without proper distance constraints, it is difficult to
state this unequivocally.

The properties of the three linear features labeled A, B, and C,
including local limiting surface brightnesses and color constraints, are
summarized in Table \ref{table:streams}. Absolute magnitudes and
luminosities are calculated assuming a distance of 11 Mpc, and are
similar to moderately-sized dwarf galaxies, although we point out that
we see no actual dwarf galaxies associated with the features themselves.
The wide range in the estimated colors is due to the low light levels of
the three features; a large uncertainty in surface brightnesses
corresponds to a much larger uncertainty in associated color. As such,
we are unable to make any definitive statements regarding the stellar
populations in these features.

We perform photometry on the three features using apertures that
include all of their visible light, subtracting off a locally-measured
background level, which produces more accurate photometric results
\citep[see][]{rudick10a}. We measure the background flux by averaging
the counts in several (4 to 6, depending on the proximity of obvious
contaminating background or foreground sources) comparably sized blank
sky regions surrounding the signal region, and use the dispersion in
the median values of the blank regions as our measure of background
uncertainty \citep[for examples, see][]{rudick10a, mihos13a}.  This
dispersion in ADU is then converted to a local limiting surface
brightness.

To determine the likelihood that these features are in any way
associated with the HI ring, we overlay HI contours from
\citet{schneider89} and \citet{md10} on our binned V mosaic for
comparison. This is shown in Figure \ref{fig:overlay}. Interestingly,
the northernmost stream, labeled A, roughly follows the northwest arc of
the HI ring. The feature may extend farther to the southwest, but is
obscured by a foreground star. Assuming an area of 1.3\arcmin \ $\times$
9.4\arcmin \ (4 $\times$ 30 kpc$^{2}$),\footnote{It should be noted that
the boxes depicted in Figures \ref{fig:mosaic} and \ref{fig:overlay} do
not correspond to the dimensions given here, and are merely for
illustrative purposes.} the total flux from the feature yields an
absolute B magnitude of $M_{B}=-12.2$, for a surface brightness of
$\mu_{B}=29.3 \pm 0.3$ mag arcsec$^{-2}$. This corresponds to a
luminosity of roughly $1.2 \times 10^{7} L_{\astrosun,B}$.

Stream B lies near M95, well away from the HI ring.  Given this
apparent separation, it may not be associated with the Ring complex at
all.  Its linear morphology and lack of any central nucleus does imply
that it has been tidally disrupted, and so it may be an object similar
to KK96 but lower in mass, possibly a satellite of M95 given its
apparent proximity to the galaxy.  Assuming an area of 1.1\arcmin
\ $\times$ 5.2\arcmin \ (3.5 $\times$ 17 kpc$^{2}$), we find
$M_{B}=-11.6$ and $\mu_{B}=29.2 \pm 0.2$, for a total luminosity of $7
\times 10^{6} L_{\astrosun,B}$.

Finally, Stream C appears to be one long linear feature, extending
from just below feature A southward almost 10\arcmin \ before
disappearing below NGC 3389. This feature is unlikely to be an
instrumental artifact, as it does not align along the rows or columns
of the CCD. It lies near a similar vertical linear feature in the HI,
although it is displaced to the east and tilted in position angle by
$\sim 1^{\circ}$. The northernmost cloud of this HI feature appears
associated with NGC 3384 in both proximity and velocity, however the
association of the two small clouds just to the south is more
ambiguous \citep{oosterloo10}. For the northern end of the optical
feature (labeled C North), sampling from an area of 1.1\arcmin
\ $\times$ 9.6\arcmin \ (3.5 $\times$ 31 kpc$^{2}$), we find
$M_{B}=-11.7$ and $\mu_{B}=29.8 \pm 1.0$. For the southern end
(labeled C South), an area of 1.3\arcmin \ $\times$ 7.0\arcmin \ (4
$\times$ 22 kpc$^{2}$) yields $M_{B}=-11.4$ and $\mu_{B}=29.9 \pm
1.0$.  Uncertainties are calculated under the assumption that the
north and south ends are pieces of one unique structure; as such,
dispersion in the ``local'' background is calculated using apertures
in blank regions around both the north and south ends. This is a
conservative approach; if the north and south ends are taken
individually, the local dispersions are much lower. For the whole
structure, if we assume an average surface brightness of
$\mu_{B}=29.9$, and a constant shape from the southern tip to the
northern tip (a total projected length of $\sim$ 200 kpc), we
calculate a total absolute magnitude of $M_{B}=-13.9$, which
corresponds to a luminosity of $5.6 \times 10^{7} L_{\astrosun,B}$.

The proximity of features A and C to similar-looking features in the
HI is intriguing. Similar streams of diffuse starlight have been found
in the Virgo cluster near M87, which are thought to be collisional in
origin \citep{mihos05}, so this may well be the case for these
features as well -- these may be the remnants of small galaxies
tidally disrupted by the group potential. In this collisional picture,
however, feature B stands out as unusual given its large distance from
the central galaxy, M105. Recent observations from the ALFALFA survey
\citep{stierwalt09} show no HI detection in its vicinity, making its
association with the M96 Group much more ambiguous.

\begin{figure*}
  \centering
  \includegraphics[scale=0.5]{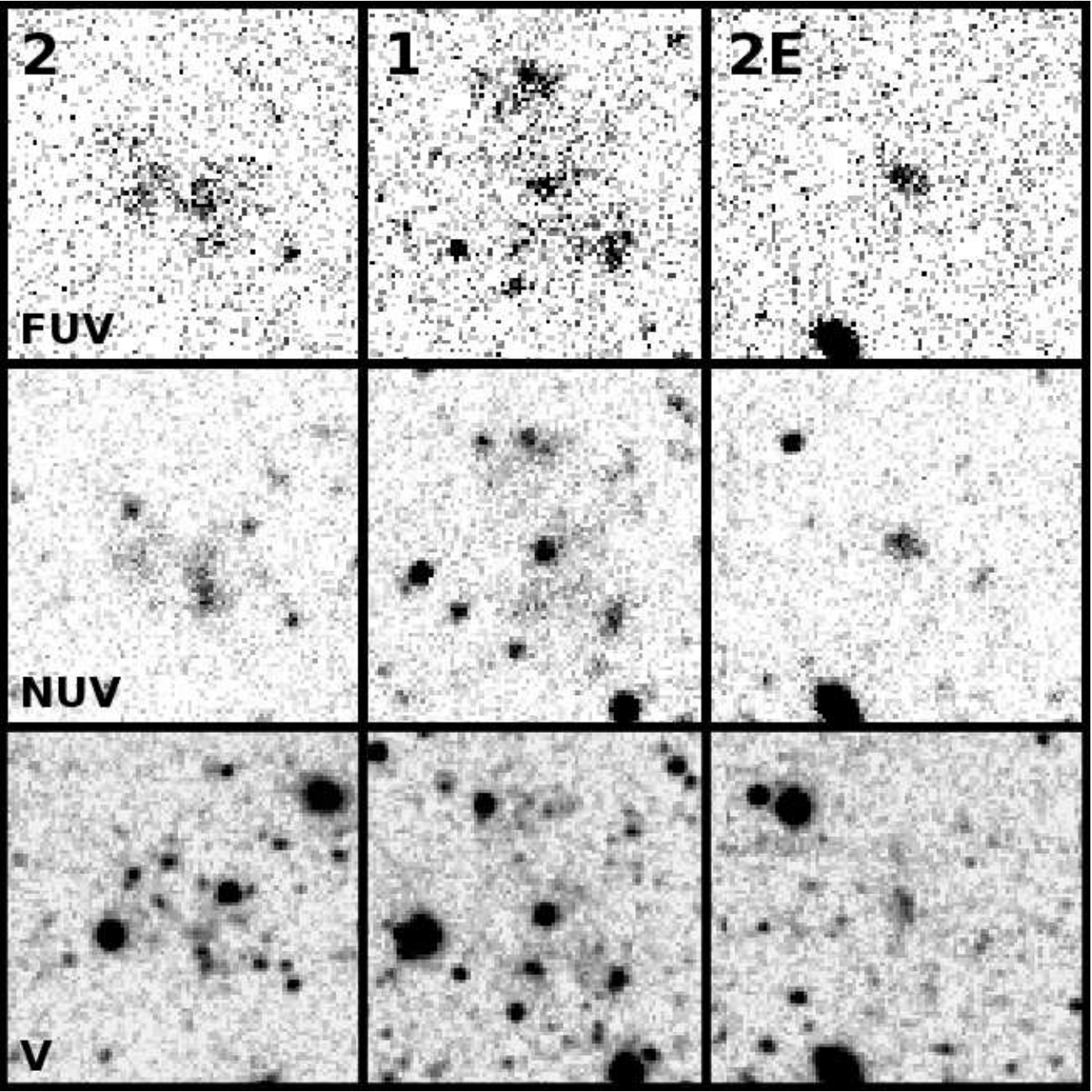}
  \caption{Comparison between GALEX FUV (top row), NUV (middle row),
    and our optical data (V band, bottom row) for the three dense HI
    regions discussed in \citep{thilker09} and \citep{md10}.  The
    region labels 2, 1, and 2E refer to the nomenclature used by
    \citet{thilker09}.  Each box is approximately 2.5\arcmin \ on a
    side (8 kpc).  Measured optical colors of the diffuse emission
    (not including point sources) in these regions are as follows.
    Region 2: $B-V = 0.2 \pm 0.2$; Region 1: $B-V = 0.0 \pm 0.1$;
    Region 2E: $B-V = 0.1 \pm 0.2$.  Uncertainties are due mainly to
    the small sizes of the regions.
       \label{fig:galex}}
\end{figure*}

Given the rough spatial correspondence between A and C and the HI, it
is useful to compare these results with some of the proposed models
for the formation of the HI ring. The collision model given by
\citet{md10}, for example, predicts a diffuse halo of starlight to
have been generated at a surface brightness $\gtrsim$29 mag
arcsec$^{-2}$. In terms of a diffuse component, again, we find nothing
up to $\mu_{B} = 30.1$ and $\mu_{V} = 29.5$; if there is such a
diffuse halo component, it must exist at levels even dimmer than
this. The model by \citet{bekki05}, on the other hand, seems to
predict no stellar component whatsoever, although it does require the
presence of a low surface brightness (LSB) disk galaxy. In their
simulation, this galaxy's gaseous disk is stripped to smaller radii,
with the stellar component left basically intact. No such small
isolated gaseous disk has been found in the HI surveys that have been
done on this group \citep{schneideretal89, stierwalt09}, and we
discover no heretofore unseen LSB disk in our imaging, despite that
the simulation by \citet{bekki05} predicts that such a remnant should
remain within 200 kpc ($1.5^{\circ}$) of the group center (see their
Figure 1).

Regarding the streamlike features themselves, given their
faintness, we can only make general guesses at their physical nature.
For example, we might assume, given the luminosities we calculate,
that these features are tidally-stripped dwarf galaxies.  Dwarf
galaxies are low mass, and hence typically have low velocity
dispersions \citep[of order 10 km/s; see e.g.][and references
  therein]{mateo98}.  After disruption, the system is no longer
gravitionally bound, and so the internal random motions should serve
to widen the remnant stream over time.  Assuming a characteristic velocity
dispersion of 10 km/s, and given their current widths of roughly 5
kpc, this would imply an interaction timescale of order 500 Myr,
shorter than the 1.2 Gyr predicted by \citet{md10}, and far shorter
than the 4 to 6.5 Gyr proposed by \citet{bekki05}.  However, without
kinematic data, it is hard to place robust constraints on formation 
scenarios for these features.

Finally, the only previously known stellar counterparts to the Ring
were the three star-forming clumps discovered by \citet{thilker09} and
later verified by \citet{md10}.  We corroborate this discovery in our
imaging as well in the form of three unresolved patches of light
coincident with -- and with morphologies very similar to -- the FUV and NUV
emission, and again with blue colors ($B-V$ between $\sim$-0.1 and
$\sim$0.4, see Figure \ref{fig:galex}).  \citet{thilker09} and
\citet{md10} came to opposite conclusions regarding the nature of this
star formation, the former finding a best-fit model to their UV colors
with a low metallicity, several hundred Myr burst, and the latter
finding that their SED was fit best by a pre-enriched, 5 Myr
instantaneous burst akin to the star-forming region NGC 5291N (within
the galaxy NGC 5291).  While we cannot constrain either model, it is
clear from our data that these star-forming regions are isolated
events within the Ring, rather than part of any extended tidal stream
of old stars, and thus represent a pure sample of stars that
have formed directly out of it.  This makes them an ideal target for
future observations in the effort to measure the metallicity of the
Ring directly.  In addition, as these regions are several kpc across,
they may represent dwarf galaxies that are in an early stage of
formation; obtaining a CMD of these stars would thus provide further
important constraints on the origin of the Leo Ring, and thus further
constraints on formation mechanisms of dwarf galaxies \citep[e.g. formation
from tidal encounters or out of primordial HI clouds;][]{lynden76,
  barnhern92, metz07, pawlowski12}.

\section{Individual Galaxies}

The streamlike features discussed in the previous section, while
intriguing, remain ambiguous in origin and nature due to poorly
constrained colors and lack of distance and velocity data.  However,
if the collisional origin hypothesis is correct, evidence of the past
encounter that produced the features may be found in the group's
galaxies themselves.  We now turn to the individual galaxies,
searching for accretion signatures and quantifying the structure and
stellar populations in their outer regions.

Here we present photometric profiles of the group's four bright
galaxies: M105, NGC 3384, M96, and M95.  Each annulus from which the
azimuthally averaged photometric values are derived takes into account
changing ellipticity and position angle, measured as a function of
radius using IRAF's \emph{ellipse} function \citep[for details,
  see][]{jedrzejewski87}.  Also for each galaxy, we measure profiles
along six constant angular-width wedges, using a constant position
angle and ellipticity derived from the last isophote fit by
\emph{ellipse}, to look for asymmetries in each galaxy that would
otherwise be washed out by azimuthal averaging.  All magnitudes and
colors are corrected for foreground Galactic extinction using the
values of $A_{V}=0.067$ and $E(B-V) = 0.02$ from \citet{schlafly11}.

\subsection{M105}

M105 (NGC 3379) is a classic elliptical galaxy, most famous for being a
keystone of the de Vaucouleurs $r^{1/4}$ photometric profile
\citep{deVau79}. Observations in HI and CO have revealed a dearth of
atomic and molecular gas \citep{bhr92,sofue93,oosterloo10}, although
there does exist a very small ($\sim$ 150 $M_{\astrosun}$) dusty disk in
the inner 13\arcsec \ \citep{pastoriza00}. The kinematics of the
galaxy's center show some peculiarities, leading to speculation about
whether it is actually a face-on triaxial S0 galaxy \citep{statler99},
although kinematics gleaned from planetary nebulae show little evidence
of rotation out to 330\arcsec \ \citep{ciardullo93, douglas07}.
\citet{schweizer92} analyzed the galaxy for evidence of recent
interaction and found none. In total, aside from the innermost regions,
the galaxy appears to be in a very relaxed, pressure-dominated state.

\begin{figure*}
  \vcenteredhbox{\includegraphics[scale=0.5]{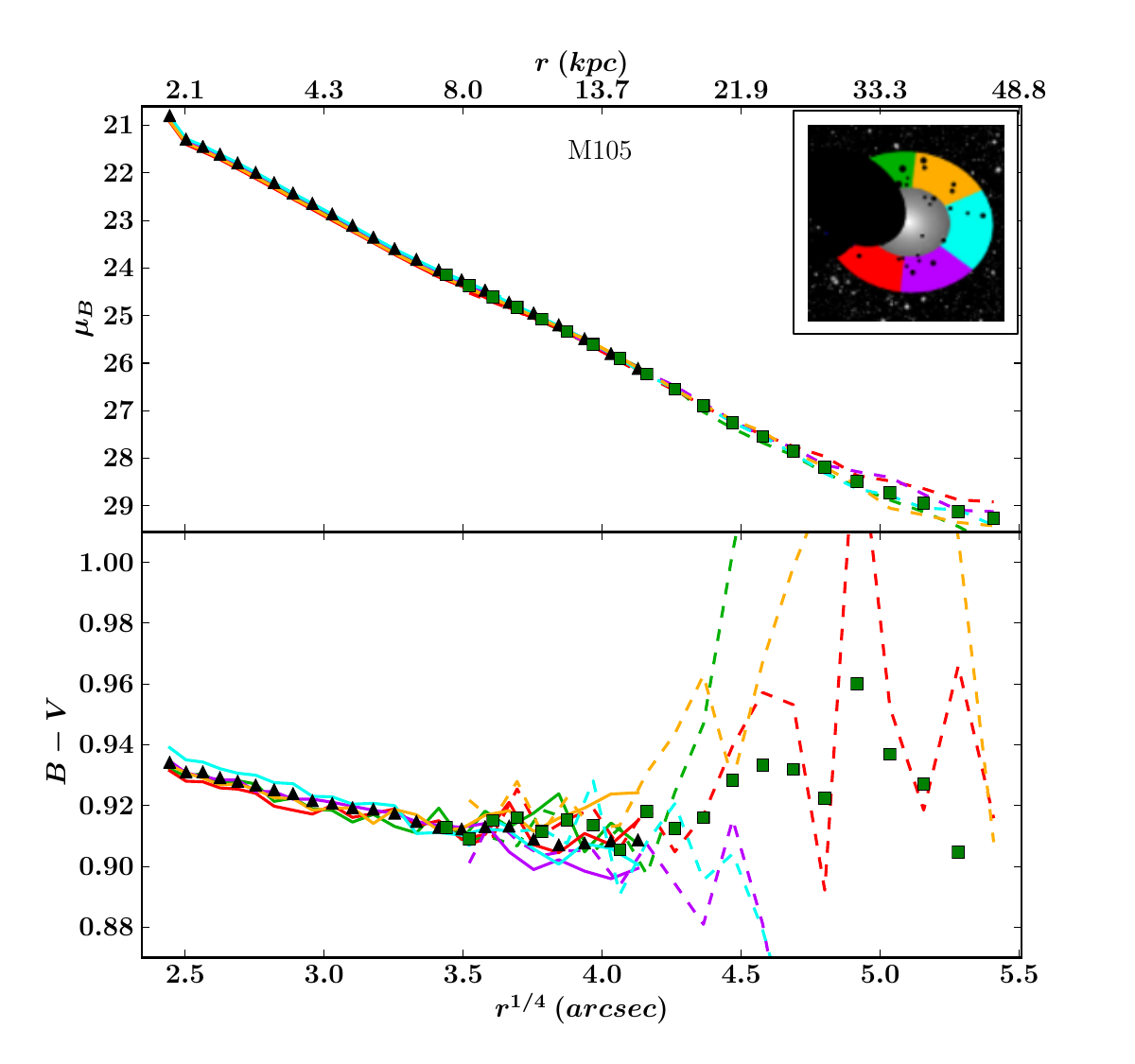}}
  \vcenteredhbox{\includegraphics[scale=0.37]{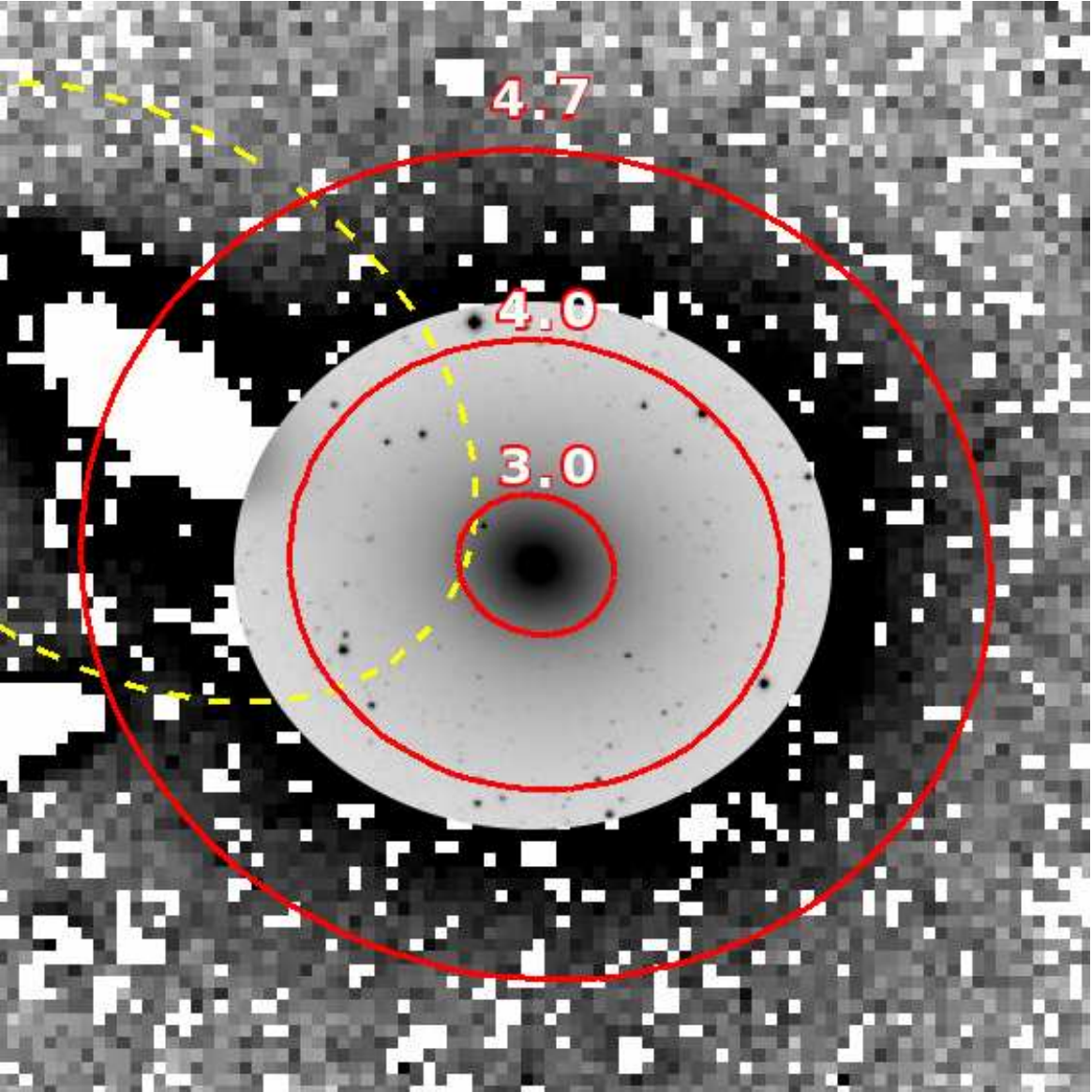}}
  \caption{{\bf Left:} B band surface brightness and $B-V$ color
    profiles for M105.  The colored lines show the profiles of
    equal-angle cuts across the galaxy, which are shown schematically
    in the upper right of the figure (solid black regions indicate
    masking).  Solid lines indicate values derived from the unbinned
    mosaic, and dashed lines indicate use of the $9\times9$ binned
    mosaic.  Likewise, black triangles are azimuthally averaged values
    in the unbinned image, and green squares are azimuthally averaged
    values in the $9\times9$ binned image.  The abscissa shows length
    of the semi-major axis in units of arcsec$^{1/4}$.  Physical scale
    is given along the top axis assuming a distance of 11 Mpc.  {\bf
      Right:} diagram of the galaxy with three labeled ellipses, for
    reference.  The values shown are in units of $r^{1/4}$ in
    arcseconds.  The unbinned image is shown inside of $r^{1/4}
    \approx 4.1$, and the $9\times9$ binned image is shown outside of
    this radius.  The yellow dotted ellipse indicates where NGC 3384
    was masked out.
       \label{fig:m105}}
\end{figure*}

Figure \ref{fig:m105} shows our photometry for this galaxy.  It can be
clearly seen that the $r^{1/4}$ model remains a good fit out to the
extent of our data ($\sim$ 850\arcsec, or 50 kpc), without significant
variations within the uncertainty.  The $B-V$ color profile shows a
slope of $\Delta(B-V)/ \Delta(\log r)=-0.04$ mag dex$^{-1}$ out to
$r^{1/4} \approx 4$, or 13.5 kpc \citep[in modest agreement
  with][]{goudfrooij94}.  Beyond $r^{1/4} \gtrsim 4.2$ (310\arcsec),
the level of uncertainty in the colors as indicated by the large
dispersion amongst the different angular profiles makes determination
of colors impossible.  Out to the 25$^{th}$ magnitude isophote, we
measure an integrated color (before reddening correction) of $B-V =
0.96$, in perfect agreement with the Third Reference Catalog
\citep[RC3][]{deVau91} value of $(B-V)_{T} = 0.96$.

\begin{figure*}
  \centering
  \includegraphics[scale=0.41]{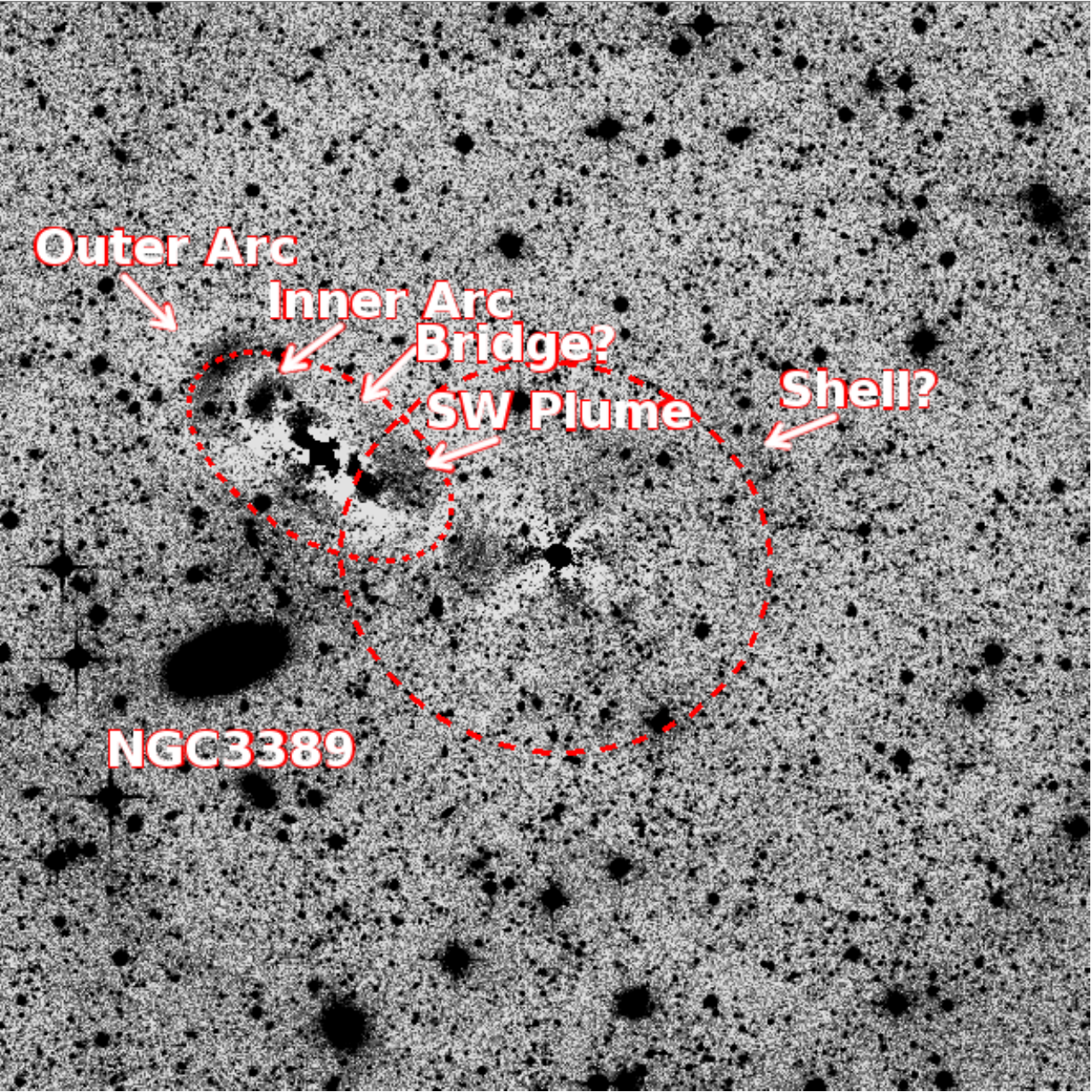}
  \includegraphics[scale=0.41]{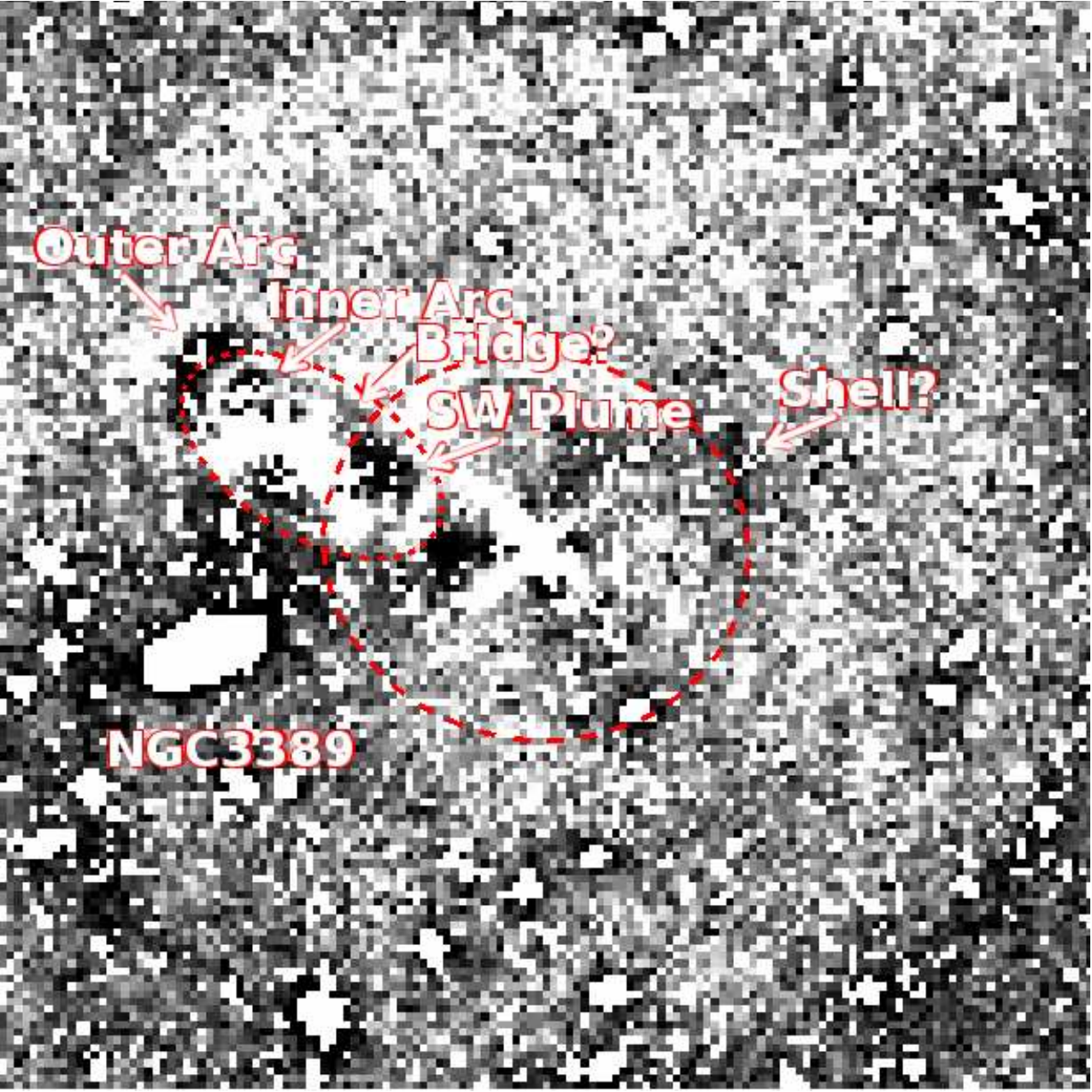}
  \caption{{\bf Left:} unbinned $B$ image of NGC 3384 and M105 after
    subtracting out models of both galaxies.  Red dotted ellipses
    correspond to the $B=26.5$ magnitude isophotes for both
    galaxies. North is up and east is to the left.  {\bf Right:} as
    left, but median binned into 9$\times$9 pixel bins.  We note
      that independent model subtractions on the V image show similar
      results.
       \label{fig:subtract}}
\end{figure*}

We see no non-axisymmetric structure in either the surface brightness
or color profiles, so to investigate more thoroughly whether M105
shows any signs of interaction, we created an elliptical model of the
galaxy using IRAF's \emph{ellipse} and \emph{bmodel} functions and
subtracted this from our image.  This is the same procedure outlined
in \citet{janowiecki10}, performed under the assumption that local
density variations will show up as residuals after subtraction of a
smooth profile.  The results are shown in Figure \ref{fig:subtract}.

The only readily apparent structure in the residual image appears near
the center of the galaxy, in the form of a pinwheel-like
structure. Such residual artifacts are commonly seen \citep[see,
  e.g.,][]{janowiecki10}, and are a sign of slight non-ellipticity in
the galaxy isophotes rather than any truly distinct structural
component. Another much fainter source appears, marginally detected, a
few arcminutes to the northwest of the galaxy's center, which may be
an extremely faint shell ($\mu_{B}\approx29.3$, measured from the
subtracted image to isolate the feature). We have verified that
  the isophotal model itself has no artifacts that would imprint such
  features, and an independent application of the technique on the V
  band image also reveals these structures.  Other than these weak
features, however, the galaxy appears to be well-modeled by smooth
elliptical isophotes with no sign of recent accretion. Consistent with
previous studies, then, M105 appears remarkable in just how
unremarkable it is.

Our data implies that if any unusual formation signatures are left to be
found in M105, they must lie in the galaxy's stellar populations.
Indeed, \citet{harris07} studied the discrete stellar populations in
M105's outer halo at a radius of $\sim$ 540\arcsec \ ($r^{1/4}=4.8$) and
found evidence for a substantial population of low metallicity stars. If
the outer halo is significantly more metal poor than the inner regions,
this should result in systematically bluer isophotes, modulo differences
in mean stellar age between the halo and inner galaxy. If anything, our
data suggest the metallicity gradient becomes more shallow, indicative
of a redder stellar population in the halo, but at the extremely low
surface brightness characteristic of the outer halo ($\mu_B > 28$), the
uncertainty in our color measurements is extremely large and makes
quantitative comparison difficult.

\subsection{NGC 3384}

NGC 3384 is a lenticular galaxy in close proximity (in projection) to
M105. Like M105, it too is nearly devoid of gas \citep{sage06,
oosterloo10} and dust \citep{tomita00}, and \citet{busarello96} found a
fairly flat color profile at around $B-V \approx 0.9$ out to 40\arcsec,
typical for S0 galaxies \citep{li11}. The galaxy also contains several
``faint fuzzies'' (Larsen et al 2001), globular cluster-like objects
with large half-light radii ($R_{eff} = 7$ - $15$ pc, versus $2$ to $3$
pc in normal GCs), which \citet{burkert05} hypothesize may have
originated in a low impact parameter galaxy collision. Previous work has
also shown the galaxy to be host to a number of other notable
peculiarities, including strong isophote twists \citep{barbon76,
busarello96}, a pseudobulge or double-bar \citep{pinkney03, erwin04,
meusinger07}, and outer boxy isophotes \citep{busarello96}. NGC 3384
thus appears much less well-settled than its neighbor M105.

\begin{figure*}
  \vcenteredhbox{\includegraphics[scale=0.5]{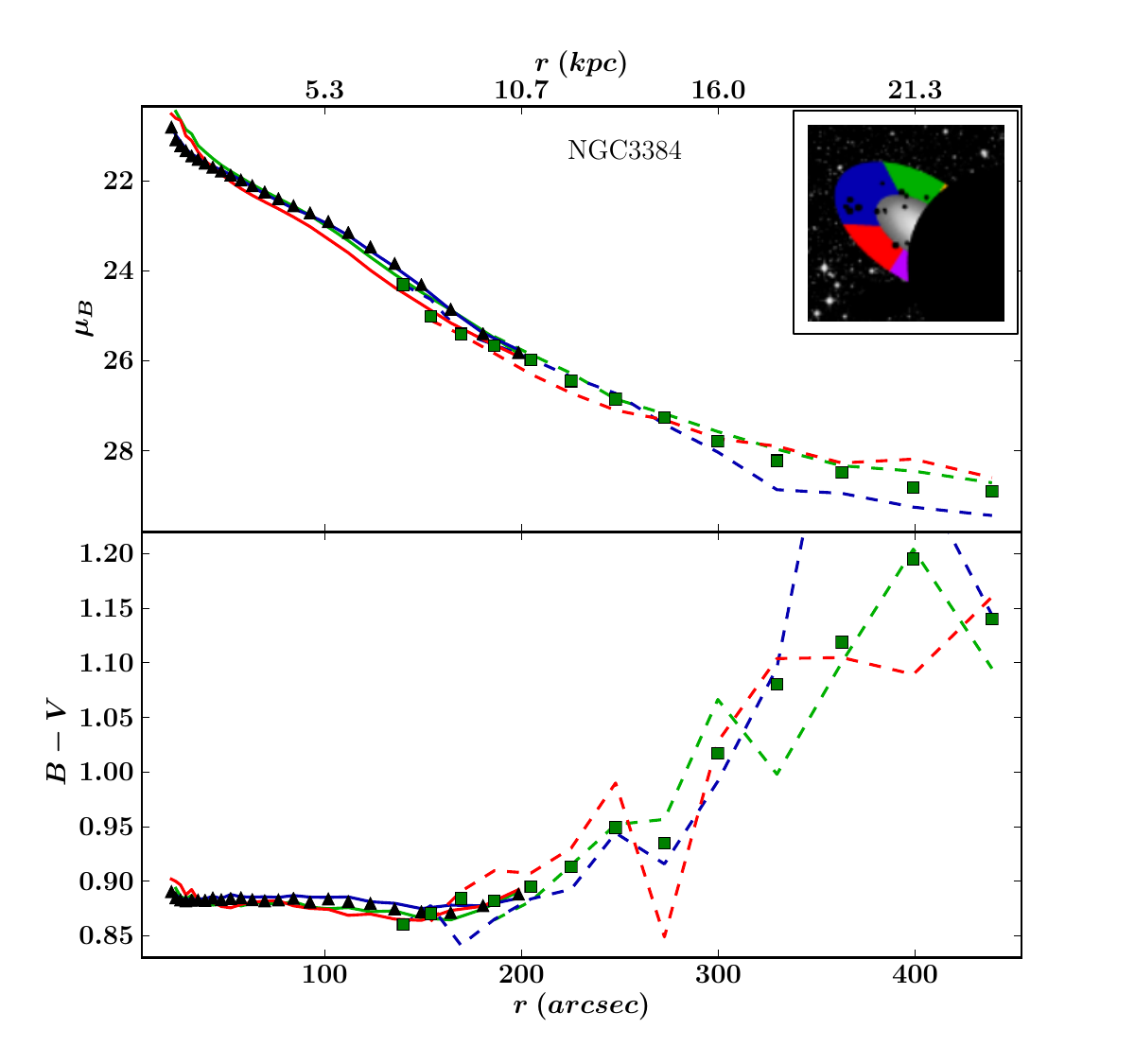}}
  \vcenteredhbox{\includegraphics[scale=0.37]{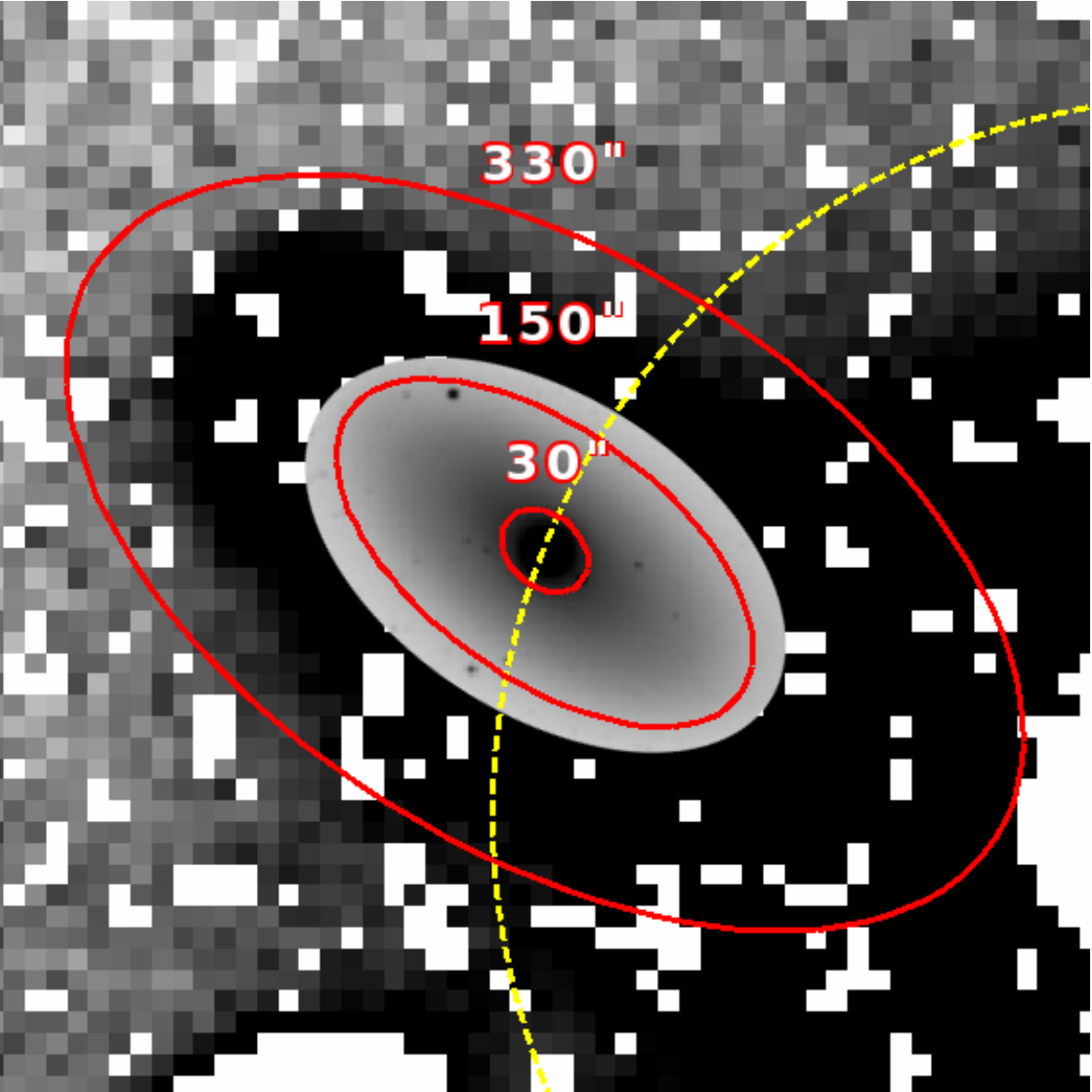}}
  \caption{As Figure \ref{fig:m105}, but for NGC 3384.  The units of
    the labeled ellipses in the right figure are now in arcseconds,
    and the yellow dotted circle represents where M105 was masked
    out.  Only full wedges are plotted.  We measure a scale length of
    36.48\arcsec, or 1.9 kpc.
       \label{fig:ngc3384}}
\end{figure*}

Figure \ref{fig:ngc3384} shows our profiles for this galaxy.  It has a
roughly exponential surface brightness profile out to
$\sim$325\arcsec; beyond this radius, the profile becomes flatter and
may simply mark the transition into the local background.  Asymmetry
is seen near the center (within 40\arcsec), resulting from the
bulge/bar complex, and from $\sim$100\arcsec - 150\arcsec, where a
hump appears in the azimuthally averaged profile.  The color profiles
appear mostly flat at $B-V = 0.88$ out to a radius of
$\sim$200\arcsec, after which they redden.  This redward color
gradient does not appear to be due to influence from the neighbor
M105; at this radius, M105's surface brightness is $\mu_{B} = 28.5$,
compared to NGC 3384's $\mu_{B} = 25.5$.  The surface brightness
profile appears to show a mild anti-truncation at this radius as well,
although the strong deviations from an exponential profile in the rest
of the disk make this difficult to quantify.  We measure an integrated
color (without extinction correction) of $B-V=0.92$, in good agreement
with the RC3 value of $(B-V)_{T}=0.93$.

To further investigate NGC 3384's outer structure, as we did for M105,
we subtracted a smooth model for NGC 3384 to search for residuals.
This is also shown in Figure \ref{fig:subtract}.  Unlike with M105,
however, we discover very clear residuals from the smooth profile in
the form of two arcs northeast of the galaxy center, as well as a
diffuse plume protruding out toward the southwest. Again, all of
  these features are seen when subtracting a model from the V image as
  well.  The northeast arcs appear much less sharp than arcs often
seen in elliptical galaxies \citep[e.g.][]{malin83}, which are thought
to be formed via minor mergers \citep{hernquinn87}.  The binned image
reveals another faint structure toward the south, as well as a
possible bridge connecting the innermost arc to the southwest plume,
which would imply more of a ring or disk shape.

The inner arc appears at a radius of $\sim$120\arcsec, and fades at
$\sim$165\arcsec.  It has a surface brightness of $\mu_{B}\approx
27.5$ with the underlying disk subtracted.  In this region, the
northeast (blue) profile from Figure \ref{fig:ngc3384} is brightest,
followed by the northwest (green), with the southeast (red and purple)
apparently following a more regular exponential disk profile.  This
behavior appears to reflect the location of the arc, though it is
interesting to note that we see no color deviations in this region.
The southwestern plume appears at roughly the same radius; however,
due to the presence of M105, we can say little about the structure of
the disk in its viscinity.  The plume has an average surface
brightness of $\mu_{B}\approx27.8$ and a much less regular morphology,
making its nature more ambiguous.  The outermost arc appears from a
radius of $\sim$200\arcsec \ to $\sim$275\arcsec, the location of the
red color gradient (and possible anti-truncation).  This arc is only
slightly fainter than the inner arc: $\mu_{B}\approx27.8$.  These
surface brightnesses correspond to absolute magnitudes of
$M_{B}\sim-11.7$ and $M_{B}\sim-12.1$ for the inner and outer arcs,
respectively, and $M_{B}\sim-12.2$ for the southwestern plume, which
implies progenitors with the luminosities of a dwarf galaxy
\citep[e.g.][]{mateo98}.  

A `faint extended luminous arc' in this galaxy, so described by
\citet{busarello96}, was photographed by David Malin in 1984 and
appears to correspond spatially to this outer arc \citep[the
  photograph is shown in Busarello et al. 1996]{malin84}.  As such, we
would suggest based on our imaging that the arc discovered by Malin is
actually part of a system of accretion features in the galaxy.  This
total system (inner arc, outer arc, and southwestern plume) has a
combined luminosity of $\sim3 \times 10^{7} L_{\astrosun}$, again
indicative of a fairly low stellar mass system.

At first glance, the surface brightness profile shown in Figure
\ref{fig:ngc3384} is suggestive of an exponential disk with one or more
breaks. However, the presence of the northwest arcs influences the major
axis profile, resulting in the hump at 100\arcsec. Instead, the profile
along the southwest wedge (shown by the red curve), away from the arcs,
more accurately follows the true underlying disk structure: a pure
exponential out to at least 16 kpc, or approximately 8 scale lengths.
Only at larger radii do we see any suggestion of a break in the profile,
in the form of a possible anti-truncation of the disk in its outermost
regions.

NGC 3384 thus provides an interesting contrast to the extremely
well-settled M105. We see that the galaxy's unusual features are not
constrained to the innermost regions -- boxy isophotes exist as far out
as 16 kpc, and the two arcs are strong evidence of disturbance. In some
ways, this galaxy is similar to the face-on S0 Arp 227, discussed in
\citet{schombert87}. That galaxy showed similar shells with much less
sharp morphology to the more standard shells often seen in elliptical
galaxies, and with colors that deviated little if at all from the
integrated color of the parent galaxy. \citet{schombert87} explained those
features as possibly arising due to an accreted and subsequently evolved
hot component from a neighboring massive galaxy (in that case, NGC 470).
In NGC 3384's case, two obvious candidates for such an interaction would
be M105 and M96, however, as shown in the previous section, M105 appears
to show little sign of any dynamical disturbance in its recent past. At
roughly 300\arcsec, NGC 3384's color is red enough ($B-V \approx 1$) to
suggest an old halo population. The surface brightness profile begins to
flatten here as well, as would be expected if an extended halo began to
dominate the profile \citep{martnav14}. However, as previously
mentioned, it is equally plausible within our uncertainty at these
levels that we are simply seeing the contaminating influence of the
scattered light in the image.

\subsection{M96}

M96 (NGC 3368) is one of two massive late-type galaxies in the group.
The galaxy has long been known to have a faint outer ring or set of
spiral arms originating at roughly 90\arcsec \
\citep[e.g.][]{schanberg73}, which we clearly resolve as part of an
outer disk with a position angle offset by more than 20$^{\circ}$ from
the high surface brightness inner disk.  These outer spiral arms
primarily have long made it (along with NGC 3384) a favored candidate
in the proposed collisional origin of the HI ring \citep{rood84,
  md10}, as this level of asymmetry is difficult to produce via
secular evolution processes.  The galaxy's connection to the Ring was
apparent from the Ring's initial discovery by \citet{schneider83} onward,
with \citet{schneider89} noting that, since most of its HI content
exists in its outer disk, M96 may actually be accreting matter from
the Ring.

\begin{figure*}
  \vcenteredhbox{\includegraphics[scale=0.5]{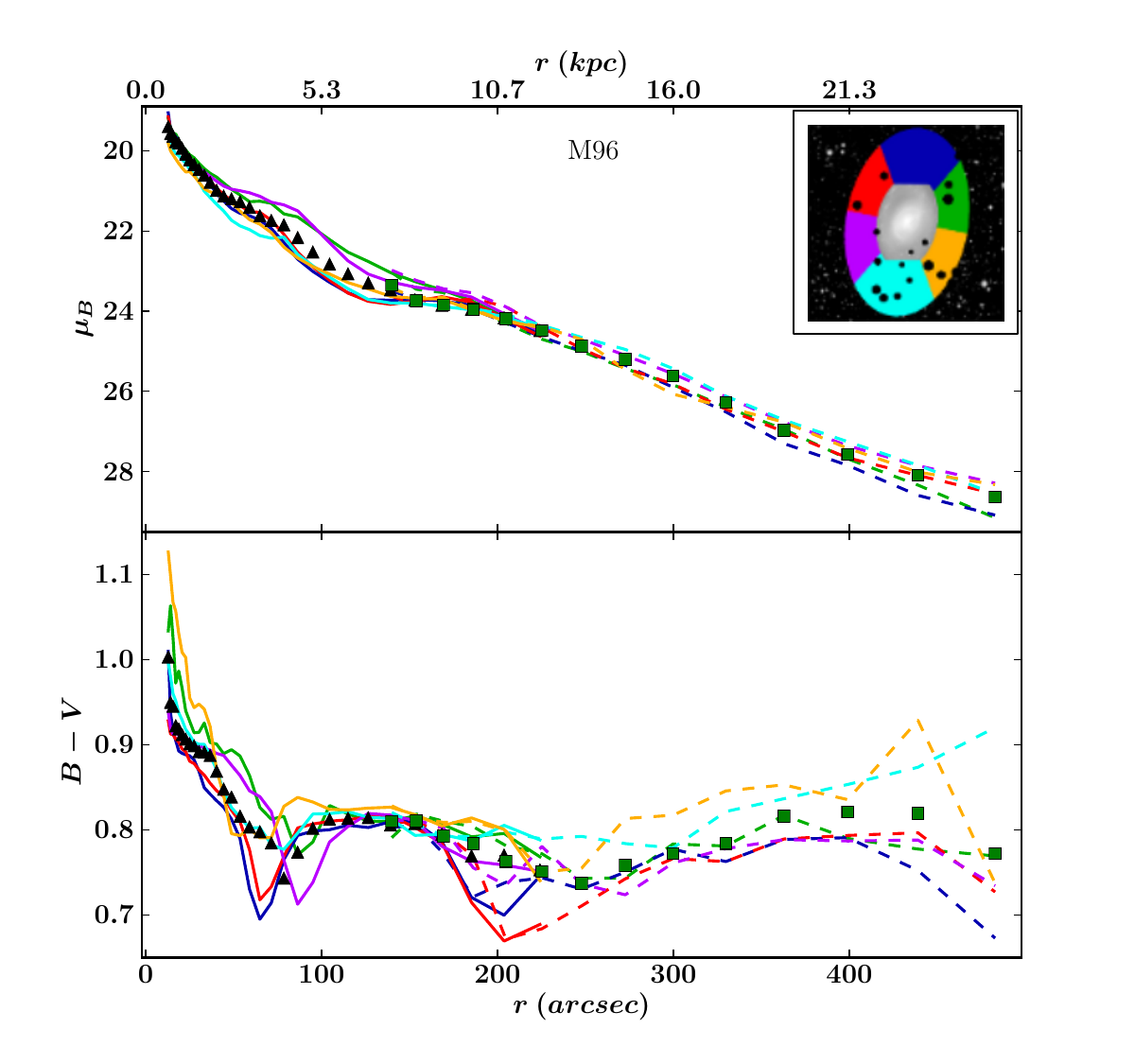}}
  \vcenteredhbox{\includegraphics[scale=0.37]{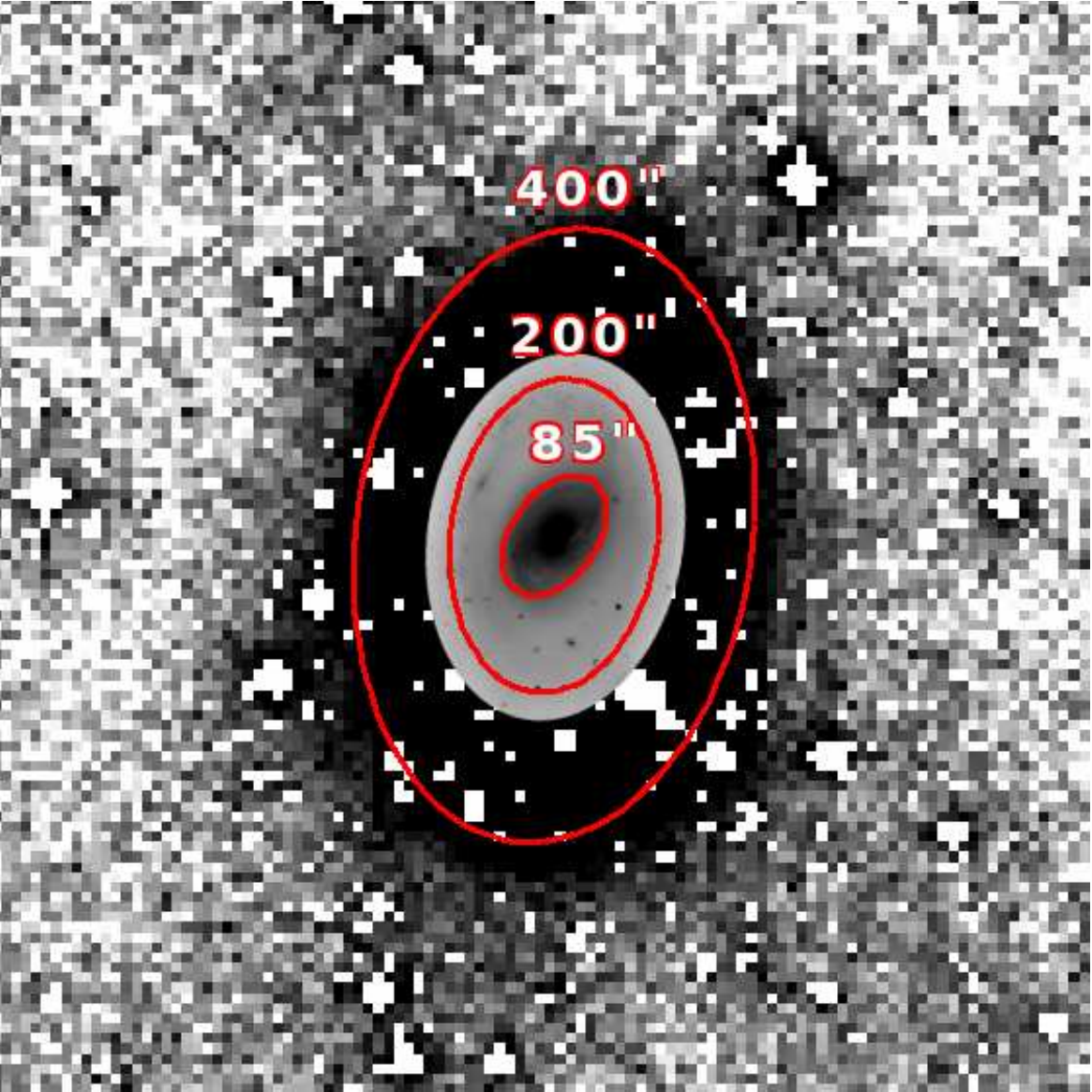}}
  \caption{As Figure \ref{fig:m105}, but for M96.  We measure a
    scale length of 61.04\arcsec, or 3.3 kpc (assuming a pure
    exponential profile).
       \label{fig:m96}}
\end{figure*}

Figure \ref{fig:m96} shows our surface brightness and color profiles
for M96. The galaxy shows a clear dip in its surface brightness
profile between 85\arcsec \ and 200\arcsec \ \citep[this `dip' is what
  has been previously described as a Type II OLR, or Outer Lindblad
  Resonance, `break' by][]{erwin08}. This appears to reflect a gap
between the inner and outer disks. It is interesting to note that
along the southeast and northwest profiles (the purple and green
wedges, respectively), we do not see the dip, and in fact these
profiles appear to have the same smooth exponential behavior as the
outer disk, implying one continuous structure rather than, for
example, an inner disk embedded in an offset outer ring. The color
profiles show an asymmetry in this `dip' region as well, bluer in the
northeast than the southwest just outside of the inner disk. Another
such asymmetric dip appears around 200\arcsec; these are most likely
caused by HII regions in the northern arm, which are not visible in
the much fainter southern arm. Beyond this radius, the color profile
appears flat with some evidence of a continued northeast/southwest
asymmetry even at extended radii, and the exponential surface
brightness profile continues unimpeded to the extent of our
measurements. Beyond 10 kpc, there is no evidence for a disk break in
the outer disk, which shows a smooth exponential profile out to 25 kpc
(7.5 scale lengths, as measured using the profile beyond
200\arcsec). Again, the integrated (uncorrected) color of $B-V=0.85$
that we measure is in good agreement with the RC3 color of
$(B-V)_{T}=0.86$.

Most ($> 60$\%) disk galaxies show breaks or truncations in their
surface brightness profiles \citep{pohlen06}, and while the inner disk
of the galaxy does show a break, the lack of a truncation in the outer
disk (beyond 10 kpc) is interesting.  If we assume, based on the
arguments of \citet{pohlen06} and \citet{erwin08}, that truncated
disks are the normal end-state for spiral galaxies, the lack of
truncation might imply that M96's disk is still being built.  For
example, mergers or accretion processes have been used to explain
anti-truncations \citep[Type III disks, e.g.,][]{penarrubia06,
  younger07} via deposition of excess light at large radius; it may
thus be that mild star-formation in the outer disk of M96 caused by
accretion from the HI ring is maintaining an exponential profile out
to very large radii.  The northeast/southwest asymmetry in our color
profiles may reflect this process: the northeast side, which shows
bluer colors at $\sim$85\arcsec \ and $\sim$200\arcsec, is the
direction from which the HI appears to be accreting
\citep{schneider89}. That these colors are only mildly bluer ($B-V =
0.7$, compared to 0.8 in the remaining angular wedges) and spatially
localized (little azimuthal mixing even at the innermost radius,
85\arcsec) would imply that such star formation, if occurring, would
be rather weak and somewhat recent, as it has not had time to
azimuthally mix. The HI bridge is also visible in more detailed HI
observations from \citet{stierwalt09}, which also indicates that M96's
HI disk is highly extended and lopsided, making the accretion
hypothesis quite plausible \citep{bournaud05}.

That said, the overall impression from the data is that M96's outer
stellar disk appears morphologically very smooth. The isophotes appear
very regular, with the centroid of the outermost isophotes drifting
south only by $\sim$15\arcsec \ (a change of only $\sim$2.5\%). If M96's
outer disk is currently being built by accretion of gas from the Ring,
it is apparently doing so in a very smooth and ordered way, despite the
one-sided nature of the accretion and the lopsidedness of the accreted
gas. A more plausible scenario may be that the accreted gas is
collecting in the outer spiral arms and forming stars in those regions:
\citet{schneider89} showed that most of the galaxy's HI content is
located within a dense ring located at roughly $\sim$200\arcsec (see
their Figure 3), the location of the outer spiral arms (see Figure
\ref{fig:m96}). This scenario does not easily explain the
single-exponential nature of the outer disk, however.

\subsection{M95}

M95 (NGC 3351) is a barred spiral galaxy hosting a well-studied,
star-forming circumnuclear ring \citep[e.g.,][]{alloin82, colina97,
elmegreen97, comeron10}, as well as a larger ring of HII regions
encircling the bar. It is the most isolated member of the group, and
appears mostly undisturbed but for the star forming activity in the
inner disk.

\begin{figure*}
  \vcenteredhbox{\includegraphics[scale=0.5]{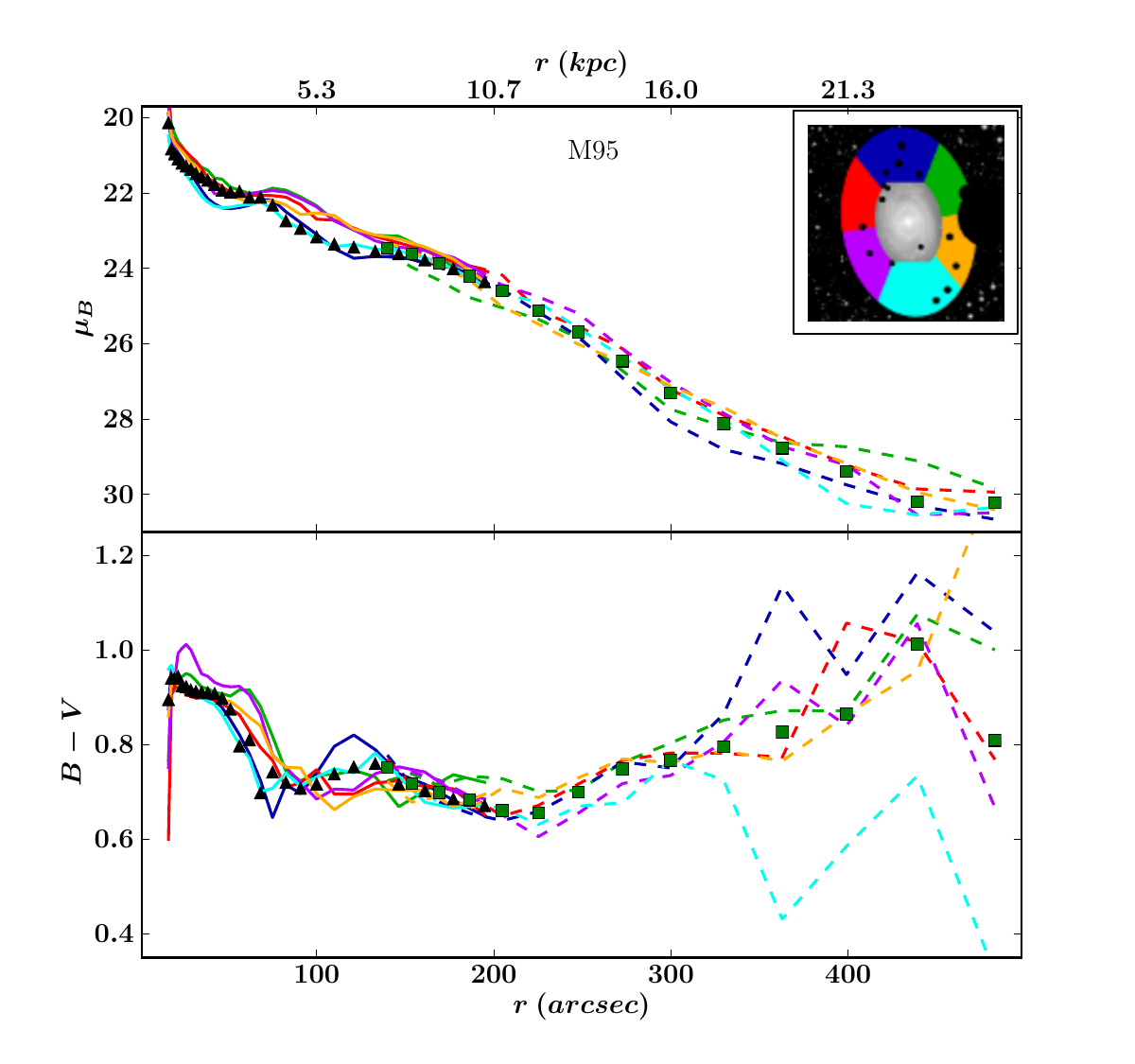}}
  \vcenteredhbox{\includegraphics[scale=0.37]{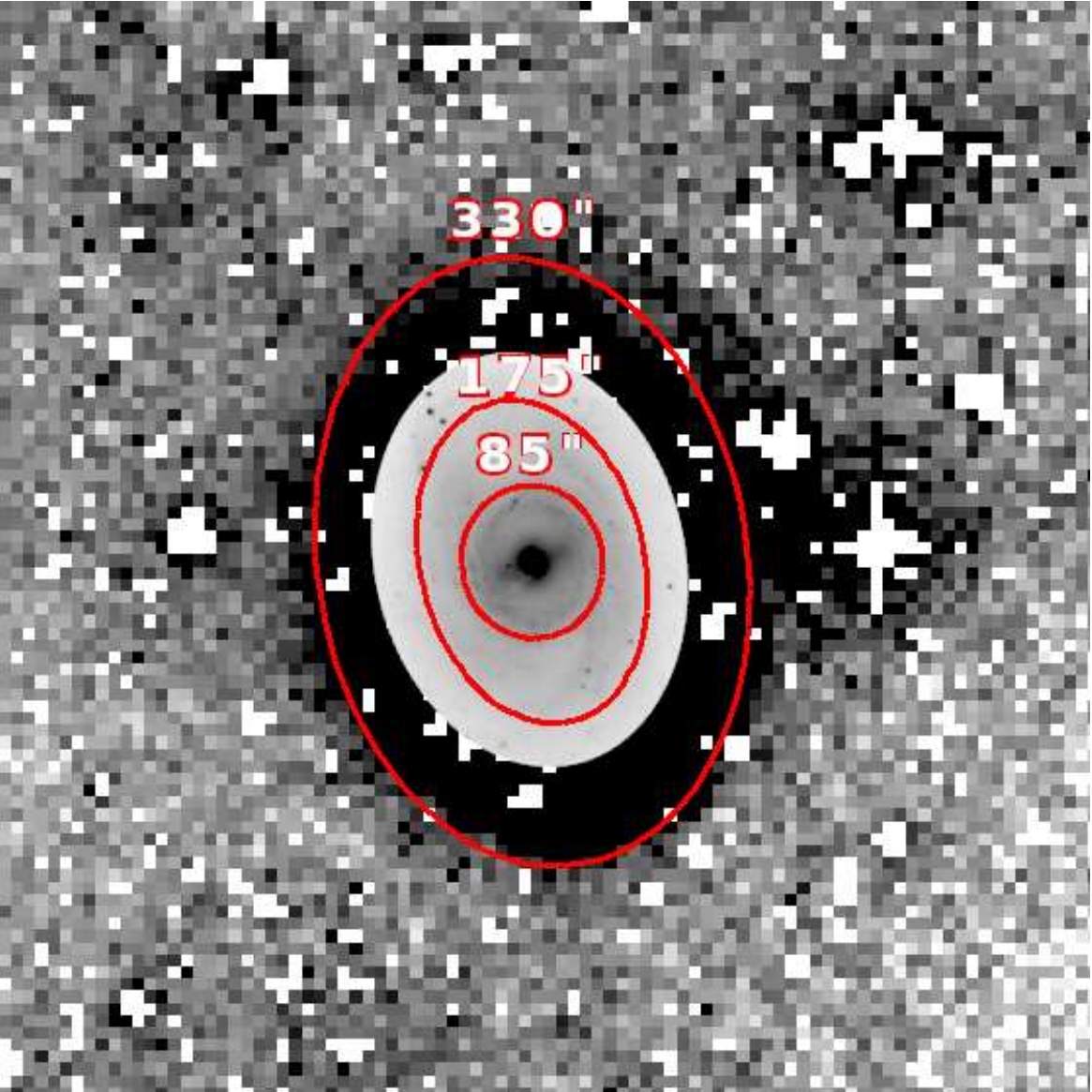}}
  \caption{As Figure \ref{fig:m105}, now for M95.  We measure an
    inner scale length of 64.26\arcsec \ (3.4 kpc), and an outer scale
    length of 42.32\arcsec \ (2.3 kpc) beyond the break at 200\arcsec.
       \label{fig:m95}}
\end{figure*}

Figure \ref{fig:m95} shows our surface brightness and color profiles
for this galaxy.  The bar and ring clearly influence the profiles
inside of 90\arcsec, with asymmetries resulting from spiral arms
between 90\arcsec \ and 130\arcsec \ \citep[again, the dip seen near
  130\arcsec \ has previously been classified as a Type II OLR
  break;][]{erwin08}.  Beyond 130\arcsec, we see a smooth exponential
disk (scale length 3.4 kpc) coupled with a mild blue color gradient,
until a break and reversal in the gradient at $\sim$200\arcsec \ (scale
length 2.3 kpc).  This broken exponential describes the disk
morphology well to the extent of our data (400\arcsec \ or 25 kpc).  The
outer isophotes of this galaxy appear quite regular, well-fit by
smooth ellipses.  No plumes or other asymmetries are visible even at
extremely low surface brightness ($\gtrsim$29 mag arcsecond$^{-2}$).
Once again, our measured integrated uncorrected color of $B-V=0.78$
agrees well with the RC3 value of $(B-V)_{T}=0.80$.

Given the presence of a distinct bar, a ring, and spiral arms in this
galaxy, as well as the regularity of the outer isophotes \citep[in
  both visible light and HI, e.g.][]{stierwalt09}, one possible
explanation for the upturn in the color profile is an age effect due
to outward radial migration of stars, driven by the non-axisymmetries
in the inner disk \citep[e.g.,][]{roskar08a, roskar08b}.  Such an
effect has been observed in other galaxies; for example, consider the
case of NGC 2403, in which \citet{williams13} discovered a flattening
of the metallicity gradient beyond $\sim$12 kpc.  Radial migration
preferentially moves stars outward \citep{roskar08a}; the farther out
a population of stars travels, the longer the travel time and hence
the older the population must be (increasing age with increasing
radius).  A sample of uniform metallicity stars, like in NGC 2403,
showing such a radial age gradient would tend to show a red color
gradient as well.  Thus a similar effect may be behind the behavior we
see in M95's outer disk.

M95 thus appears to have no strong signatures of recent accretion in
its outer disk. While its inner regions are tumultuous --
\citet{sersic67} classified it as a `hotspot' galaxy -- most of the
star formation and other activity is well-described by resonances
with, for example, the stellar bar \citep{devereux92, helfer03}. Given
the galaxy's membership in the M96 Group, any interactions with other
group members must have been extremely weak or have happened in the
distant past, long enough ago that the outer disk has had time to
recover. In \S 3, we suggested that Stream B (20\arcmin or 65 kpc away
from M95, see Figure \ref{fig:overlay}), rather than being associated
with the Leo HI ring, may have resulted from the tidal disruption of a
satellite galaxy interacting with M95. If so, the encounter appears to
have made little impact on M95's outer disk.

\section{Discussion}

\subsection{The origin of the Leo Ring}

In the context of the formation of the Leo Ring, arguably the most
intruiging aspect of the M96 Group, the lack of either a diffuse stellar
counterpart or strong interaction signatures in any of the group's four
most massive galaxies is extremely puzzling. The imprint of strong
interactions on the outskirts of galaxies (in the form of tidal features
or distorted isophotes) should remain over the long dynamical timescales
of the outer disks, yet we see no sign of such features in our imaging.
This makes it hard to envision dynamical scenarios that invoke {\it
recent} interactions to produce the Leo Ring. Even with our deep
imaging, however, conclusive evidence favoring any particular formation
scenario remains lacking. We describe below in more detail the connection
between the observed structure of the Leo Group galaxies and the various
formation scenarios for the Leo Ring.

\subsubsection{The case of M96 and NGC 3384: the collisional origin}

M96 and NGC 3384 (the two galaxies commonly proposed to have generated
the Ring) demonstrate the difficulty in reconciling our results with
collisional origin models for the Ring; while both galaxies show signs
of disturbance (the arcs found in NGC 3384, and the position angle
offset between the inner and outer disk of M96), it is unclear whether
or not these could be the remnants of an interaction of the type
proposed by \citet{rood84} and \citet{md10}. For instance, the type of
morphology seen in M96 has occasionally been referred to as an `oval
distortion' and may actually be a result of secular evolution
\citep{kormendy04}. The galaxy M94 (NGC 4736), for comparison, has a
very similar morphology to M96, yet appears to have no nearby companions
of significant mass \citep{trujillo09}.

However, the mildness of the interaction signatures we do find in the
outer disks must constrain the timescale of any interaction model, as
clearly any significant perturbation caused by an interaction has by
now been erased.  The simulation by \citet{md10} required 1.2 Gyr for
the Ring to achieve its current size, which is roughly the orbital
period of M96's extreme outer disk \citep[based on the gas velocities
  of][]{schneider89}. Yet M96's outer disk, the region most loosely
bound and hence most easily perturbed, appears mostly smooth, with the
exception of the slight southern skew shown in Figure
\ref{fig:m96}. It is difficult to imagine how the galaxy's outer disk
could settle so rapidly (in one orbital period) after such a strong
encounter as envisioned by \citet{md10}. To displace such a massive
amount of gas into the Leo Ring while only leaving behind such mild
signatures in M96's outer stellar disk would require a high
interaction velocity \citep{spitzer51}, while the group's low measured
velocity dispersion \citep{pierce85, stierwalt09} would seem to imply
a more tightly bound system.  The timescale of 6.5 Gyr proposed by
\citet{bekki05} may be more realistic in this regard, but again, the
expected signature their model would leave behind (a gas-deprived, low
surface brightness galaxy) does not appear to be present.  The
original scenario proposed by \citet{rood84} appears even less likely
due to its much shorter timescale (500 Myr).

Alternatively, rather than an encounter disturbing any pre-existing
outer disk, it is conceivable that M96's outer disk was created by the
encounter.  The apparent accretion of HI onto M96 from the Ring is
interesting in this context: recall the bifurcation in the color
profile, with slightly bluer colors on the side nearest the Ring
($B-V=0.7$) than on the farthest side ($B-V=0.8$).  While these colors
are similar to the more quiescent, early type disk galaxies
\citep[Sa-Sab and S0a, respectively; see:][]{roberts94}, these are
colors averaged over wedges; the northern spiral arm itself shows a
number of large patches with colors of roughly $B-V=0.4$, while the
southern arm is nearly devoid of such regions.  This is reflected in
the GALEX FUV data as well \citep{martin05}, in which both arms are
clearly visible at 200\arcsec, but the southern is visibly fainter
than the northern.  Measuring the FUV flux in the northern arm from
the GALEX image and adopting the conversion to star formation rate
given by \citet{kennicutt98}, we find a total star formation rate of
0.002 \sfrone over an area of roughly 45 kpc$^{2}$, whereas in the
southern arm we find only $5 \times 10^{-4}$ \sfrone ~over an area of
20 kpc$^{2}$.  This thus implies star formation surface densities of
$4\times 10^{-5}$ \sfrtwo ~in the north and $2 \times 10^{-5}$ \sfrtwo
~in the south, a factor of two lower.  That said, these derived SFRs do
not take into account dust attenuation, which can be a significant
factor in deriving SFRs from UV flux due to scattering and absorption
effects; \citet{buat05}, for example, found a mean FUV attenuation of
1.6 magnitudes in their galaxy sample (which was dominated by
late-type galaxies).  A more detailed analysis of the SFR in M96
should of course take these effects into account.

Nonetheless, this is still apparently fairly low-level star formation
that could conceivably be induced by slow accretion from the HI ring,
which itself totals only $10^{9} M_{\astrosun}$.  We do also see star
formation continuing within the southern arm, so if we assume that
this is leftover from the initial accretion where the HI ring meets
the galaxy's disk, we can place a lower limit on the timescale of
accretion of one orbital time at the 10 kpc radius where we see the
spiral arms, which is roughly 400 Myr.  This does not seem to pose a
problem for the 1.2 Gyr interaction timescale proposed by
\citep{md10}, or even the 500 Myr timescale proposed by \citep{rood84}
if we assume gas began accreting immediately after the time of the
interaction.  However, it is also true that the current star formation
rate is not high enough to build the outer disk in a reasonable time
frame; at this rate, to produce the total disk luminosity beyond
200\arcsec \ would require longer than a Hubble time.  As such, the
accretion timescale given here cannot differentiate between a
collisional scenario, in which M96 began accreting HI shortly after it
had been ejected from NGC 3384, and a non-collisional scenario, in
which M96 began accreting HI from a pre-existing cloud 400 Myr ago or
earlier by simply passing near it.  Given all of these ambiguities, it
is important simply to reiterate that we find no \emph{obvious}
indications in our data for this galaxy that any particular
collisional model is the correct one.

Regarding NGC 3384, the embedded arcs discussed in section 4.2 are
unusually broad, but again, they do not appear to be signatures of a
major interaction. As stated previously, the symmetric morphology of
these features strongly resembles shell systems or caustics formed via
minor mergers. Such artifacts can be formed from mergers involving low
luminosity disk or spheroidal companions \citep{quinn84, hernquinn87},
or simply from accretion of material from a passing galaxy
\citep{hernquinn87}. While it can be difficult to constrain the type
of the progenitor from the morphology alone, they reflect better the
models of dwarf elliptical disruption shown by \citet{hernquinn88}. To
conserve phase-space volume, shells formed this way should evolve to
become sharper over time \citep{hernquinn88}, so the thickness of the
shells may thus imply a rather recent merger (or mergers). Minor
mergers are also capable of heating disks \citep{walker96} and leaving
behind long-lasting (multiple orbit) warps \citep{quinn93} in outer
regions.  Also, thick-disk formation models using minor mergers tend to
show increasing boxiness with decreasing surface brightness
\citep{villalobos08}, so the presence of such signatures in NGC 3384
again does not immediately imply a major, gas-stripping interaction. The
nearest extended tidal feature to the galaxy, feature C in Figures
\ref{fig:mosaic} and \ref{fig:overlay}, is not at all aligned with the
galaxy's disk, making it unlikely that it constitutes stripped
material from NGC 3384. The only HI clearly associated with the galaxy
is also ambiguous in nature and origin \citep{oosterloo10}. So, once
again, we find no particularly clear evidence in favor of any given
interaction model based on either of these galaxies' properties.

\subsubsection{The case of M95 and M105: passively evolving systems}

The other large spiral in the group, M95, shows no signs of recent
interactions; instead, its appearance is completely consistent with
mild secular evolution. The only suggestion of a past interaction is
the possible connection with the feature B seen in Figures
\ref{fig:mosaic} and \ref{fig:overlay}. Given that this galaxy lies on
the outskirts of the group, its undisturbed morphology may simply be
another manifestation of the mechanisms responsible for the
morphology-density relation \citep{oemler74, dressler80}. It is
somewhat intriguing, however, that M95 is apparently not host to any
satellites capable of visibly disturbing the isophotes out to 25 kpc
(given their smoothness at this radius).  Regardless, the galaxy
appears to have no connection to the HI ring.

This leaves the elliptical galaxy M105, sitting at the center of the
group and surrounded almost perfectly by the HI ring.  Yet as our deep
imaging shows, it is extremely smooth and relaxed. We find only a
marginal detection of any interaction signature, in the form of the
faint shell-like structure marked in Figure
\ref{fig:subtract}. Considering the plethora of low mass members of
this group discovered in the HI \citep{stierwalt09}, it is truly
remarkable just how unperturbed this galaxy is. If the Leo Ring was
generated via strong interactions within the group, it is perplexing
that the central galaxy shows no evidence of such an encounter.

Yet given the uncertain nature of the interaction model for forming
the Leo Ring, it may be that a more reasonable explanation for the
Ring's origin lies with M105 itself. Indeed, extended complexes of
neutral hydrogen surrounding normal early type galaxies are not rare
\citep[e.g.,][]{vangorkum86, franx94, appleton90, schiminovich00,
  oosterloo07, oosterloo10, serra12}. While the original ``primordial
origin'' concept for the Leo Ring argued that the HI is a pristine
remnant from the early universe out of which the group members formed,
modern galaxy formation models have ellipticals forming
hierarchically, through major mergers of smaller objects. This process
can be quite messy and, if gas-rich galaxies are involved, can eject a
significant amount of HI out of the remnant in the form of extended
tidal tails. If this gas could settle into an extended ring, this
could place the formation of the Leo Ring at an intermediate age --
likely many Gyr ago, given the relaxed state of M105 and lack of any
observed tidal features around it. However, this idea still suffers
from the problem of longevity \citep[how the Ring can be stable for
  gigayears given the short group crossing
  times:][]{schneider85}. Additionally, it suffers from an angular
momentum argument: gas in merger simulations tends mostly to shock and
lose angular momentum, sinking to sub-kpc scales and initiating
starbursts, with only a fraction remaining in intermediate scale
\citep[similar to Cen A:][]{vangorkom90} disks
\citep[e.g.][]{mihos96}. It would thus seem extremely difficult to
create a 200 kpc diameter ring of low column-density gas in this way.

\subsection{The lack of intragroup light}

Our deep imaging shows no extended diffuse intragroup light within the
M96 Group, down to a limit of $\mu_{B}$ = 30 mag arcsec$^{-2}$, save
for a few small low surface brightness streams reminiscent of tidally
disrupted dwarfs. These objects aside, the lack of a significant
extended IGL component in this group is puzzling.  For example,
\citet{sommer06} predicted via group simulations that anywhere from
12\% to 45\% of the light found in groups should exist in an IGL
component by the present day, with a spatially patchy distribution. We
do not find anything similar to this in the M96 group. Excluding any
undetected diffuse component, the amount of IGL we find constitutes an
essentially negligible fraction of the total group light
($<0.01$\%). \citet{sommer06} did find that the amount of IGL
increases as the group evolves, with the so-called ``fossil groups''
\citep[as defined by][]{donghia05} having suffered the most
processing. The M96 Group does not qualify as a fossil group by the
\citet{donghia05} criteria,\footnote{namely, that the second brightest
  galaxy is at least 2 magnitudes fainter than the brightest galaxy;
  in the M96 Group, all four galaxies are within one magnitude of each
  other.} so it may simply be that the group is not dynamically
evolved enough to have generated a substantial amount of intragroup
light.  However, this conclusion is somewhat at odds with M105's very
relaxed state, which suggests that that galaxy, at least, is a well
evolved system. An alternative model comes from simulations by
\citet{kapferer10}, which argue that the fraction of intragroup
stellar mass (and hence IGL) should \emph{decrease} over time due to
low frequency of interactions; new stars form in the galaxies over
time, but few new stars are ejected to contribute to the IGL. Even so,
these models predict that the intragroup stars still contribute from
3\% to 30\% of the total group stellar mass, which remains higher than
what we find.  In general, it seems that the M96 Group simply does not
easily fall into either of the paradigms described by
\citet{kapferer10} and \citet{sommer06}.

Comparing observations to simulations is not always straight-forward,
however. For example, the stream-like features found in our imaging
data, if indeed part of the M96 Group, have luminosities that fall
well below the mass resolution in either the \citet{kapferer10} or
\citet{sommer06} models. Furthermore, the brightest galaxies in the
\citet{sommer06} simulations, around which the IGL is centered, are
apparently much brighter than M105. Those model galaxies show surface
brightnesses of $\mu_{B} \lesssim 26.5$ at a radius of 39 kpc (see
their Figure 3), while M105 is already below $\mu_{B}=28$ at this
radius.  This raises the possibility that the M96 Group simply hosts a
correspondingly dimmer IGL halo, however if this is the case, it would
be at such low surface brightness that it would not contribute more
than a few percent of the total group light, again suggesting a much
lower IGL fraction than that predicted by the simulations.

Of course, from the observational perspective, quantitative
comparisons to simulations or even to other data depends strongly on
how one defines IGL in the first place \citep[good demonstrations of
  this are shown in][]{kapferer10, rudick10b}, as well as how one
defines a ``group'' of galaxies. If, for example, we assume a
projected area of the M96 Group of $\sim0.13$ Mpc$^{2}$ (using a rough
radius of 200 kpc; see captions of Figures \ref{fig:mosaic} and
\ref{fig:overlay}) and use our constraints on the observed IGL (an
upper limit of $\mu_{B}=30.1$; the light of the three detected streams
is negligible), the upper limit on the diffuse starlight of this group
comes to $\sim$4\% of total. If we assume instead that the M96 Group
and the M66 Group are a single, larger group \citep[as has been
  suggested due to their low mutual velocity dispersion and similar
  distances; see][]{stierwalt09}, and that diffuse starlight extends
uniformly throughout a circular area encompassing both subgroups (a
radius of about 4$^{\circ}$, or 800 kpc), the IGL fraction increases
to 20\%, although it is important to reiterate that these are absolute
upper limits: we in fact make no such detection). Including the
extended disks or halos of the group galaxies would artificially add
light to the IGL, hence the often rather large ranges of such
quantities found in the literature.

Given this, it is perhaps more reasonable to compare to other
observations of group environments on a more qualitative
level. Previous observations of the loose M101 Group showed a similar
lack of an extended, diffuse IGL component down to a limiting surface
brightness of $\mu_{B}=29.5$ \citep{mihos13a}. However, observations
of compact groups by \citet{darocha05} and \citet{darocha08} show
smooth envelopes of diffuse light surrounding most of the galaxy
groups in their sample. Observations of more massive, even denser
environments such as galaxy clusters can show even larger fractions of
diffuse intergalactic light \citep[e.g.][]{adami05}, so it may simply
be that local density plays the largest role.

IHL (intrahalo light) in general is thought to be mostly composed of
stars that have been gravitationally stripped from their host galaxies
\citep[e.g.][]{napolitano03, murante04, rudick06, purcell07}.  It is
thus reasonable to assume that the two most influential factors in the
generation of IHL in any given environment are the strength of
interactions and the frequency of interactions.  These two factors in
turn depend on two physical quantities: mass and density of the
cluster or group \citep[mass being essentially a proxy for the number
  of galaxies;][]{dressler80}.  For example, the Coma Cluster is one
of the densest and most massive clusters in the nearby universe, and
the diffuse component of this cluster is dense and bright enough to
have been identified in early photographic imaging by
\citet{zwicky51}, although substructure was not detected until much
later \citep{gregg98, adami05}.  The Virgo Cluster, by contrast, is
less massive and dense than Coma, and hosts a fainter ICL
component, again with notable substructure \citep{mihos05,
  janowiecki10, rudick10a}.  On the low-mass, high-density end, large
fractions of IGL (20\%--50\%) have been found via imaging compact
groups of galaxies \citep{nishiura00, white03}, but with considerable
scatter from group to group \citep{darocha05, darocha08}, and some
compact groups having little or no IGL \citep{aguerri06, darocha08}.
With fewer galaxies in play, interactions should be less frequent;
this implies that the actual fraction of stellar mass that is
liberated from host galaxies will depend more strongly on the details
of each individual interaction \citep[e.g. relative masses,
  inclination angles, and velocities of the interacting
  galaxies;][]{toomre72, negroponte83}, thus giving rise to a higher
dispersion in IGL properties.

The M96 Group may thus fit into this picture by occupying the low-mass,
low-density regime of galaxy environments. Measuring the total mass of
the M96 Group is not trivial, since the entire Leo~I Group has
considerable substructure spatially and kinematically
\citep[e.g.][]{stierwalt09}. A simple application of the virial theorem
using the the velocity dispersions and harmonic mean radii from
\citet{stierwalt09} yields masses for the M96 Group in the range of 2--6
$\times 10^{12} M_{\odot}$. Using the calibration between velocity
dispersion and group mass given by \citet[their equation 6]{yang07}
gives a rough halo mass of 8 $\times 10^{12} M_{\odot}$. While it is
highly unlikely that the M96 Group is virialized, it seems clear that
the M96 Group is less massive than the mass range of previous loose
group studies \citep[typically 10$^{13} M_{\odot}$ -- 10$^{15}
M_{\odot}$;][]{castro03, feldmeier04b, durrell04}. These more massive
loose groups have low IGL fractions as well (2\%), yet still seem to
contain a higher fraction than that detected in the M96 Group, or the
still less massive M101 Group \citep{mihos13a}. We may thus be seeing a
continuation of this pattern: lower density and lower masses yield
lower IHL fractions.

In this context, the Leo Ring may thus serve as a vital clue regarding
the types of dynamics seen in low-mass, loose groups such as M96.
This is a rare system in the local universe, in terms of the size and
mass of the HI complex.  A somewhat similar structure resides in the
NGC 5291 complex located 58 Mpc away at the edge of the Abell 3574
cluster, but this HI ring is much more massive and contains a number
of large, obviously star-forming clouds \citep{malphrus97, boquien07}.
As discussed in Section 3, star formation in the Leo Ring is much more
subtle: it seems to only amount to the three kpc-scale clumps
discovered by \citep{thilker09}.  \citet{bot09} did make a tentative
discovery of dust within the Ring, near (but not precisely coincident
with) the region labeled 1 in \citep{thilker09}.

These discoveries are beneficial, as obtaining a metallicity for the
ring would be instrumental in constraining the Ring's origin. Again,
however, the analyses of these star-forming clumps by
\citet{thilker09} and \citet{md10} produced conflicting results
($1/50$ solar metallicity in the former and $1/2$ solar in the
latter), and \citet{bot09} considered their dust detection too
uncertain to make any definitive conclusion. The Ring also appears
devoid of obvious HII regions \citep{donahue95} and planetary nebulae
\citep{castro03}. As such, these very localized star forming regions
may be the extent of star formation within the Leo Ring. Other avenues
for studying star formation and the stellar content of the Leo Ring
include the possibility of deep space-based imaging to search for
discrete stellar populations in the Ring, and even deeper ground based
searches for diffuse H$\alpha$ indicative of additional on-going star
formation throughout the Ring. Such studies are clearly needed to
understand the formation and evolution of this structure, and the
galaxy group surrounding it.

\section{Summary}

We have performed deep ($\mu_{B,{\rm lim}} \approx 30$ mag arcsec$^{-2}$)
imaging of the M96 Group in order to search for low surface brightness
intragroup light, as well as to study the morphologies and stellar
properties of the outer disks and halos of galaxies in the group. We
find no diffuse stellar counterpart to the group's 200 kpc diameter HI
ring down to $\mu_{B} = 30.1$, but identify three extremely faint
($\mu_{B} \sim 29.2$ to $29.9$) streamlike features apparently
associated with the group. Two of these features may be directly
associated with the HI ring, as they show some amount of coincidence
with similar-looking features in the distribution of neutral hydrogen. 

We constructed surface brightness and color profiles (both azimuthally
averaged and in discrete angular wedges) for the group's four most
massive galaxies -- M105, NGC 3384, M96, and M95 -- out to $\sim$25
kpc in each disk galaxy, and out to $\sim$50 kpc in the elliptical
M105. We find no evidence of recent disturbance in either M105 or M95,
though the latter shows a clear disk truncation and redward color
gradient beginning around 12 kpc, possibly due to the effects of
radial stellar migration. We also find two arcs embedded in NGC 3384's
disk, as well as a strong redward gradient in the outer isophotes
reaching old halo-like colors ($B-V \approx 1$), which appears to be
associated with the outermost arc. M96 shows mild asymmetry in its
extreme outer disk, as well as a color asymmetry apparently reflecting
mild star formation likely induced by accretion from the HI ring; no
disk break is found in the outer disk out to the extent of our data,
possibly indicating that this component of M96 is still being built.

The lack of strong tidal disturbances in the outskirts of the group
galaxies, coupled with the absence of significant starlight associated
with the Leo Ring provides some tension for models which rely on recent
strong encounters between group galaxies to form the HI ring. The
connection between the Ring and the individual galaxies remains unclear;
however, further study of the small diffuse structures detected in our
imaging may allow for further testing of the various theories proposed
to explain the origin of the Ring. Finally, unless the M96 Group has
been caught in the very early stages of evolution -- unlikely, given the
relaxed state of its central elliptical, M105 -- the extremely low
fraction of intragroup light we measure also places constraints on the
ability of groups of this mass and density to act as ``pre-processors''
for the more commonly seen intracluster light found in massive galaxy
clusters.

\section{Acknowledgments}

This work has been supported by the National Science Foundation via
grants 1108964 (J.~C.~M.) and 0807873 (J.~J.~F.) and Research
Corporation grant 7732 (J.~J.~F.). We also thank Tom Oosterloo for the
use of his WRST HI data in the construction of Figure~\ref{fig:overlay},
and Pat Durrell for many useful discussions. This research made use of
NumPy, SciPy \citep{oliphant07}, and Matplotlib \citep{hunter07}.
Figures \ref{fig:m105} through \ref{fig:m95} made use of Min-Su Shin's
(University of Michigan) publicly available code img\_scale.py
\footnote{http://dept.astro.lsa.umich.edu/$\sim$msshin/science/code/Python\_fits\_image/}.

{\it Facility:}
\facility{CWRU:Schmidt} - The Burrell Schmidt of the Warner and Swasey
Observatory, Case Western Reserve University

\end{document}